\documentclass[%
 reprint,
 amsmath,amssymb,
 aps,
]{revtex4-1}

\usepackage{graphicx}
\usepackage{dcolumn}
\usepackage{bm}
\usepackage{hyperref}

\usepackage{xcolor}
\newcommand{\red}[1]{}
\renewcommand{\red}[1]{{\color{red}{#1}}}

\newcommand{\blue}[1]{}
\renewcommand{\blue}[1]{{\color{blue}{#1}}}

\newcommand{\magenta}[1]{}
\renewcommand{\magenta}[1]{{\color{magenta}{#1}}}

\newcommand{\orange}[1]{}
\renewcommand{\orange}[1]{{\color{orange}{#1}}}

\begin{document}


\title{Benchmarking a nonequilibrium approach to photon emission in relativistic heavy-ion collisions}%

\author{Anna Sch\"afer$^{1,2}$}
\author{Juan M. Torres-Rincon$^3$}
\author{Jonas Rothermel$^{4,1,2}$}
\author{Niklas Ehlert$^1$}
\author{Charles Gale$^5$}
\author{Hannah Elfner$^{4,2,1}$}

\address{$ˆ1$ Frankfurt Institute for Advanced Studies (FIAS), Ruth-Moufang-Stra{\ss}e 1, 60438 Frankfurt am Main}
\address{$ˆ2$ Institut f\"ur Theoretische Physik, Johann Wolfgang Goethe-Universit\"at, Max-von-Laue-Strasse 1, 60438 Frankfurt am Main, Germany}
\address{$ˆ3$ Department of Physics \& Astronomy, Stony Brook University, Stony Brook, New York 11794, USA}
\address{$ˆ4$ GSI Helmholtzzentrum f\"ur Schwerionenforschung, Planckstrasse 1, 64291 Darmstadt, Germany}
\address{$ˆ5$ Department of Physics, McGill University, 3600 University Street, Montreal, QC, H3A 2T8, Canada}

\date{\today}

\begin{abstract}
In this work, the production of photons through binary scattering processes is investigated for equilibrated hadronic systems. More precisely, a nonequilibrium hadronic transport approach to describe relativistic heavy-ion collisions is benchmarked with respect to photon emission. Cross sections for photon production in $\pi + \rho \to \pi + \gamma$ and $\pi + \pi \to \rho + \gamma$ scattering processes are derived from an effective chiral field theory and implemented into the hadronic transport approach, SMASH (Simulating Many Accelerated Strongly-interacting Hadrons). The implementation is verified by systematically comparing the thermal photon rate to theoretical expectations. Further, the impact of form factors is discussed, scattering processes mediated by $\omega$ mesons are found to contribute significantly to the total photon production. Several comparisons of the yielded photon rates are performed: to parametrizations of the very same rates as used in hydrodynamic simulations, to previous works relying on different cross sections for the production of direct photons from the hadronic stage, and to partonic rates. Finally, the impact of considering the finite width of the $\rho$ meson is investigated, where a significant enhancement of photon production in the low-energy region is observed. This benchmark is the first step toward a consistent treatment of photon emission in hybrid hydrodynamics+transport approaches and toward a genuine dynamical description. \\
\end{abstract}

\pacs{Valid PACS appear here}
\maketitle



\section{\label{Intro}Introduction}
Photons are valuable and direct probes of the strongly interacting medium created in heavy-ion collisions. Not only do they escape the fireball scarcely affected, they are also produced in all stages of a heavy-ion collision, thus providing a time-integrated picture of the evolution. A number of heavy-ion experiments are currently being carried out at different facilities covering a wide range of collision energies. The major goal of these efforts is the understanding of strongly interacting matter at extreme temperatures and densities.  \\
  The measured direct photon emission (i.e. all photons excluding decay photons from long-lived hadronic decays)  in heavy-ion collisions currently lacks a complete theoretical explanation. Most prominently, the  photon yield and elliptic flow measured at RHIC (Relativistic Heavy-Ion Collider)\citep{PHENIX:2010_photonYieldAuAu200GeV, PHENIX:2012_photonFlow200GeV, STAR:2014_QM2014_photonYield} and LHC (Large Hadron Collider)\citep{ALICE:2019_2.76TeV_photons, ALICE:2015_photonYield_2.76TeV,ALICE:2012_PhotonFlow_2.76TeV, Wilde_ALICE:2013_QM2012_PhotonYield_PbPb2.76TeV_pp_7TeV} can still not be described simultaneously within any theoretical calculation \citep{PHENIX:2015_200GeVAuAu, Paquet_SotA_Photon_Paper, Shen:2016_TheoryOverviewEM, EMMI_RRTF}. Previous effort has gone into dissolving the tension between theory and experiment, either by focusing on initial state phenomena \citep{McLerran_Schenke:2016_Modification_quark_gluon_distributions, Bzdak:2012_MagneticField_Eccentricity, Mohanty:2011_InitialState+2+1D_Hydro, Liu:2009_QGP_extractInitialCondition, Liu:2015_delayedQGP}, bulk and medium effects \citep{Muller:2013_Flow_with_magnetic_field, Basar:2012_Conformal_Anomaly}, thermal emission from the plasma \citep{Monnai:2014_lateQuarkFormation, Greif:2016_BAMPS_photons, AMY_Rates:2001, PHSD:2016_photons, Liu:2015_delayedQGP, Dion:2011_viscousPhotons} or hadronic emission from the late stages \citep{PHSD:2016_photons, Linnyk:2013_CentralityDep_Photons_PHSD, UrQMD:2010_photons,  Liu_Rapp:2007_Bremsstrahlung, Heffernan_Rapp_omega_parametrization}. In addition, attempts to couple photon production from the thermal plasma and from the hadron gas have been made. Therein, the space-time evolution of the system is usually modeled by means of hydrodynamics and photon emission is calculated by folding the temperature evolution with photon rates \citep{Gale:2014_Photons_Dileptons_QGP+HM, Holopainen:2011_QGP_HG_Hydro}. Alternatively, transport approaches are being used to describe photon production based on microscopic cross sections for the partonic stage \citep{Greif:2016_BAMPS_photons}, the late stages \citep{UrQMD:2010_photons}, or the entire evolution of the system \citep{PHSD:2016_photons}. There are hints that photons originating from the late and dilute rescattering phase of heavy-ion collisions might contribute significantly to the generation of direct photon elliptic flow \citep{ Paquet:2017QM, Shen:2016_TheoryOverviewEM, Shen:2014_PhotonThermometer, HeesGaleRapp:2011_PhotonsBNL, VanHees_Rapp:2015_fireballModel}. In the late stage of the evolution the momentum asymmetries of the medium are fully developed and photons inherit the elliptic flow from hadrons. \\
This work focuses on the mesonic production of photons in the late, dilute stages. In previous works within the transport approaches UrQMD (Ultra-relativistic Quantum Molecular Dynamics) \citep{UrQMD:2010_photons} and PHSD (Parton-Hadron String Dynamics) \citep{PHSD:2016_photons, Linnyk:2013_CentralityDep_Photons_PHSD} mesonic photon production is implemented relying on cross sections from \cite{Kapusta:1992gv}. In SMASH (Simulating Many Accelerated Strongly-interacting Hadrons) \citep{Weil:2016zrk, SMASH_github} cross sections based on the effective field theory described in \citep{Turbide_PhD,Turbide:2003si}, suitable to describe a larger number of photon production processes, are incorporated. The current work provides details of the calculated cross sections and a benchmark of the corresponding thermal rates compared to analytic expectations. The importance of processes involving the $\omega$ meson are confirmed, as found in \citep{Rapp_omega}. These processes can not be described within the framework provided in \citep{Kapusta:1992gv}. In addition, the more complete framework described in \citep{Turbide_PhD,Turbide:2003si} is successfully applied in hydrodynamic simulations. Therefore, a hybrid approach employing SMASH as an afterburner to describe the entire evolution of a heavy-ion collision should rely on the same input, for consistency. \\
In the following, the applied effective chiral field theory is first described in Sec. \ref{Theory} before the cross section are determined in Sec. \ref{XSections}. In Sec. \ref{Model} the hadronic transport approach SMASH is introduced with particular emphasis on the implementation of photon production in Sec. \ref{Photons_SMASH}. This implementation is validated in Sec. \ref{Rates} by comparing the thermal photon rates from SMASH to theoretical expectations. The effect of introducing form factors is further studied in Sec. \ref{FF}. In continuation, the resulting photon rates from SMASH, derived within the framework described in \citep{Turbide:2003si}, are compared to parametrizations of the very same rates in Sec. \ref{Turbide_Comp}. In Sec. \ref{Kapusta_Sec}, they are further compared to another set of hadronic photon rates as derived in  \citep{Kapusta:1992gv}, and to partonic AMY photon rates \citep{AMY_Rates:2001, AMY_Rates:2001_2}. Finally, the previously described framework is extended to broad $\rho$ mesons in Sec. \ref{broad}. A summary and an outlook are further presented in Sec. \ref{Outlook}. In addition, the differential cross sections of the presented framework, the thermal photon rates and further details regarding the extension to broad $\rho$ mesons and the implementation in SMASH are provided in Appendices \ref{app:level1}-\ref{app:param}.


\section{\label{XSec_Sec}Photon Cross Sections from an Effective Chiral Field Theory}

\subsection{\label{Theory}Theoretical framework}
  The theoretical framework in which the cross sections are calculated, is described in detail in \cite{Turbide:2003si, Turbide_PhD}. Thus, only the main features are covered in the following, and the interested reader is referred to the original publication. \\
  The underlying theory can be classified as a chiral effective field theory with mesonic degrees of freedom. It follows from a massive Yang-Mills approach \cite{PhysRev.96.191}, capable of describing pseudoscalar, vector and axial vector mesons and the photon. The corresponding Lagrangian reads as
  \begin{widetext}
  \begin{align}
  	\mathcal{L} = & \ \dfrac{1}{8} \ F_{\pi}^2 \ Tr\left(D_{\mu}UD^{\mu}
  	U^{\dagger}\right) \ + \ \dfrac{1}{8} \ F_{\pi}^2 \  Tr\left(M\left(U
  	+U^{\dagger}-2\right)\right) - \ \dfrac{1}{2} \ Tr \left(F_{\mu\nu}
  	^{L}F^{L \mu\nu} \ + \ F_{\mu\nu}^{R} \ F^{R \mu\nu} \right) \notag
    	+ \\&
   	 \ m_0^2 \ Tr\left(A_{\mu}^LA^{L\mu} + A_{\mu}^RA^{R\mu} \right) \
  	+ \ \gamma \ Tr \left(F_{\mu\nu}^{L}U F^{R \mu\nu} U^{\dagger}
  	\right) \ \notag  - \ i \xi \ Tr \left(D_{\mu}UD_{\nu}U^{\dagger}
  	F^{L\mu\nu} + D_{\mu}U^{\dagger}D_{\nu}UF^{R\mu\nu} \right) - \\&
  	\ \dfrac{2em_V^2}{\tilde{g}} B_{\mu} \ Tr \Bigl(Q
  	\tilde{V}^{\mu}\Bigr) \ - \ \dfrac{1}{4}
  	\left(\partial_{\mu} B^{\nu} - \partial_{\nu} B^{\mu}\right)^2 + \
  	\dfrac{2 e^2 m_0^2}{g_0^2} B_{\mu}B^{\mu} Tr\left(Q^2\right)
  	+ \ g_{VV\phi} \ \varepsilon_{\mu\nu\alpha\beta} \
  	Tr \Big[\partial^{\mu} V^{\nu} \partial^{\alpha} V^{\beta} \phi \Big] ,
  	\label{Lagrangian}
  \end{align}
  \end{widetext}
  where
    \begin{align}
    \begin{aligned} 
    U &= \mathrm{exp}\left(\dfrac{2i}{F_{\pi}} \ \phi \right) \\[+0.1cm]
  	\phi &= \begin{pmatrix}
  	\frac{\pi^0}{\sqrt{2}} + \frac{\eta_8}{\sqrt{6}} & \pi^+ & K^+ \\
  	\pi^- & \frac{\eta_8}{\sqrt{6}} - \frac{\pi^0}{\sqrt{2}} & K^0 \\
  	K^- & \bar{K}^0 & -\frac{2}{\sqrt{6}} \ \eta_8
  	\end{pmatrix} \\[+0.1cm] 
  	A_{\mu}^L &= \frac{1}{2} \ (V_{\mu} + A_{\mu})\\[+0.2cm] 
  	A_{\mu}^R &= \frac{1}{2} \ (V_{\mu} - A_{\mu}) \\[+0.2cm] 
  	D_{\mu} &= \partial_{\mu} - i g_0 A_{\mu}^L + i g_0 A_{\mu}^R \\[+0.2cm] 
  	F_{\mu\nu}^{L,R} &= \partial_{\mu} A_{\nu}^{L,R} - \partial_{\nu} A_{\mu}^{L,R} - i g_0 [A_{\mu}			^{L,R},A_{\nu}^{L,R}] \\[+0.2cm]
  	M &= \dfrac{2}{3} \left(m_K^2 + \dfrac{1}{2} m_{\pi}^2 \right) - \dfrac{2}{\sqrt{3}} \left( m_K^2 			- m_{\pi}^2 \right) \lambda_8 . \\[+0.2cm]
  \end{aligned}
  \end{align}
  In the above, $\phi$, $V_{\mu}$ and $A_{\mu}$ denote the pseudoscalar, vector and axial vector meson fields, respectively. $F_{\pi}$ is the pion decay constant and $\lambda_i$ are the Gell-Mann matrices. The remaining parameters are chosen such that they correspond to set (II) in the categorization made in \cite{Turbide:2003si}. For the sake of reproducibility, the values of these parameters as used in our computation can be found in Appendix \ref{app:param}. Note that $\rho$ mesons are treated as stable particles, neglecting their finite width. In Sec. \ref{broad}, an attempt is made, to apply the derived framework to a system where the width of the $\rho$ meson is explicitly taken into consideration. \\
  The above described theoretical framework is further extended by applying hadronic dipole form factors of the kind
  \begin{align}
  	\hat{F}(t) = \left( \dfrac{2 \Lambda ^2}{2 \Lambda^2 - \bar{t}_{\pi / \omega} (E_{\gamma})} 				\right)^2 ,
  	\label{hadronic_dipole_FF}
  \end{align}
  where $\Lambda = 1 $ GeV and $\bar{t}$ can be parametrized as a function of the photon energy ($E$), for $\pi$ and $\omega$ meson exchange. In \cite{Turbide_PhD}, these parametrizations read
  \begin{align}
  \begin{aligned}
  	\bar{t}_{\pi} = \ & 34.5096 \ \mathrm{GeV}^{-0.737} \ E^{0.737} \ -   \\& 67.557 \ \mathrm{GeV}^{-0.7584}  \ E^{0.7584} \ +  \\& 32.858 \ \mathrm{GeV}^{-0.7806} \ E^{0.7806}, \label{FF_pi} \end{aligned}
  	 \\[+0.4cm]
  	\begin{aligned}
  	\bar{t}_{\omega} = & -61.595 \ \mathrm{GeV}^{-0.9979} \ E^{0.9979} \ +  \\& \ 28.592 \ \mathrm{GeV}^{-1.1579} \ E^{1.1579} \ +  \\& \ 37.738 \ \mathrm{GeV}^{-0.9317} \ E^{0.9317} \ -  \\& \ 5.282 \ \mathrm{GeV}^{-1.3686} \ E^{1.3686}.
  	\label{FF_omega}
  	\end{aligned}
  \end{align}
  Note that, for simplicity, form factors are applied directly to the final cross sections, rather than to each specific vertex individually. This is possible since the parametrizations, and therefore also the form factors defined in Eq.~(\ref{hadronic_dipole_FF}), depend on the photon energy only, and do not rely on knowledge about the kinematic details of the underlying scattering process.

\subsection{\label{XSections}Photon cross sections}
\begin{figure}
  	\includegraphics[width=0.45\textwidth]{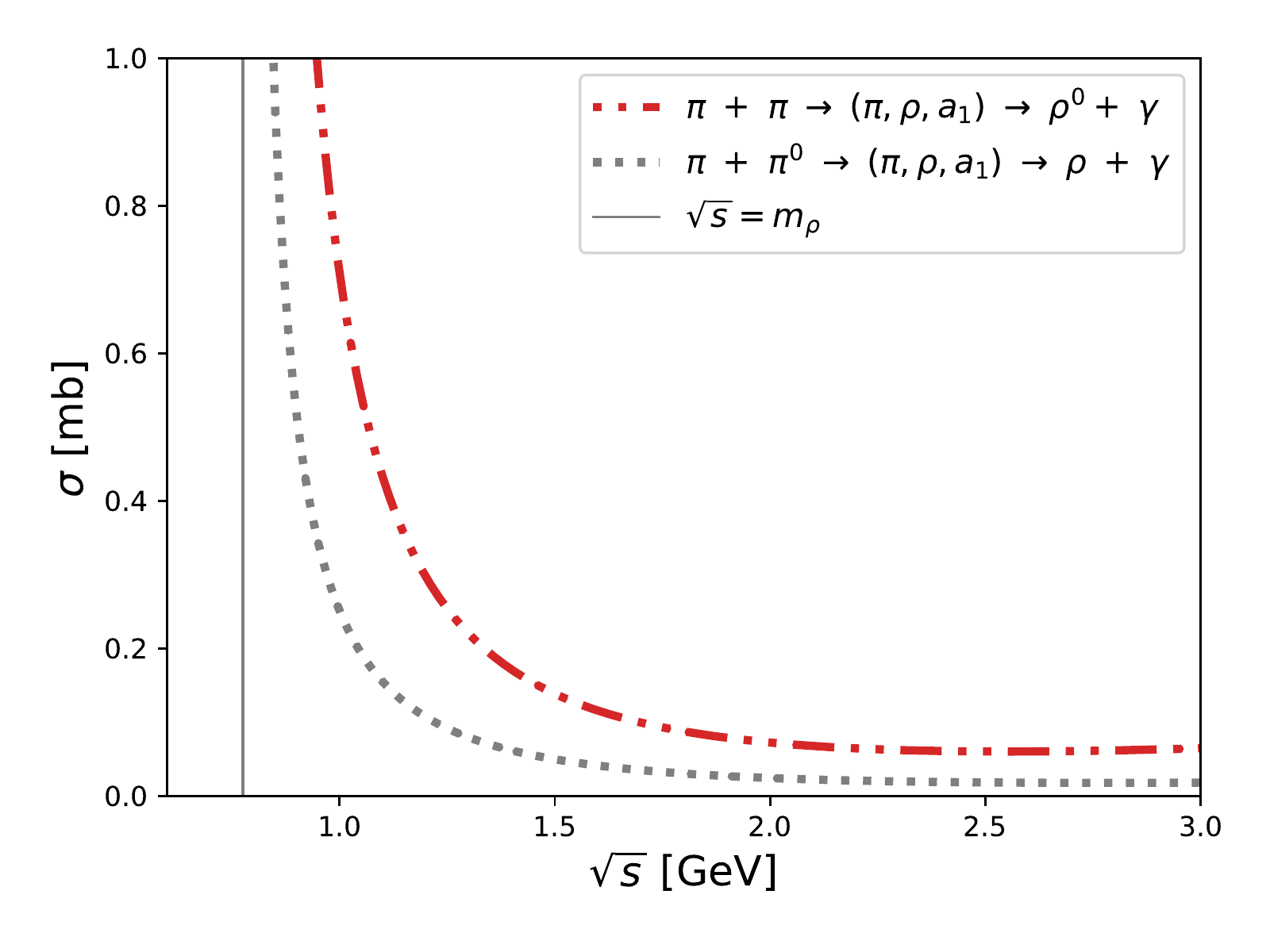}\\[-0.3cm]
  	\includegraphics[width=0.45\textwidth]{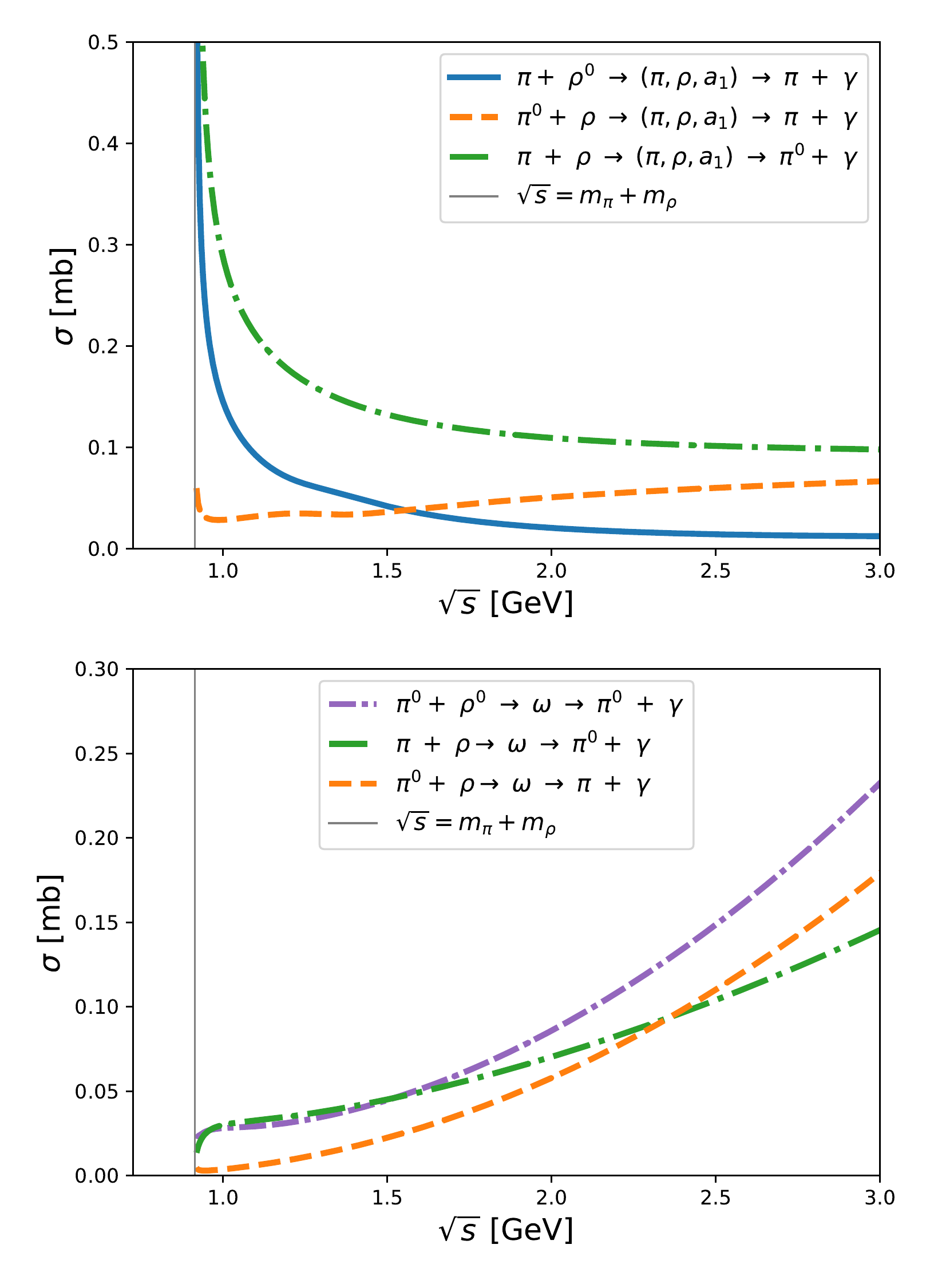}
  	\caption{Cross sections of ($\pi, \rho, a_1$)-mediated $\pi + \pi  \rightarrow \rho+ \gamma$ processes (upper panel), ($\pi, \rho, a_1$)-mediated (center panel) and ($\omega$)-mediated (lower panel) $\pi + \rho \rightarrow \pi+ \gamma$ processes as a function of $\sqrt{s}$. }
  	\label{Sigmas}
  \end{figure}

  There are 8 different photon production channels that are currently implemented in the SMASH transport approach.
  Following the logic in \cite{Turbide:2003si}, they are categorized into processes mediated by either $(\pi, \rho, a_1)$ mesons or the $\omega$ meson. The considered processes are: \\
  \begin{subequations}
  \begin{align}
  	\pi^{\pm} + \pi^{\mp} \rightarrow (\pi,&\rho,a_1) \rightarrow \rho^0 + \gamma \label{C21}\\
  	\pi^{\pm} + \pi^0 \rightarrow (\pi,&\rho,a_1) \rightarrow \rho^{\pm} + \gamma	\label{C22} \\[+0.5cm]
  	\pi^{\pm} + \rho^0 \rightarrow (\pi,&\rho,a_1) \rightarrow \pi^{\pm} + \gamma \label{C11}\\
  	\pi^0 + \rho^{\pm} \rightarrow (\pi,&\rho,a_1) \rightarrow \pi^{\pm} + \gamma \label{C12}\\
  	\pi^{\pm} + \rho^{\mp} \rightarrow (\pi,&\rho,a_1) \rightarrow \pi^0 + \gamma \label{C13}\\[+0.5cm]
  	\pi^0 + \rho^0 \rightarrow \ &\omega \rightarrow \pi^0 + \gamma \label{C14}\\
  	\pi^{\pm} + \rho^{\mp} \rightarrow \ &\omega \rightarrow \pi^0 + \gamma \label{C15}\\
  	\pi^0 + \rho^{\pm} \rightarrow \ &\omega \rightarrow \pi^{\pm} + \gamma \label{C16}
  \end{align}
  \end{subequations}\\
  Note that these processes involve solely pions and $\rho$ mesons as initial or final state particles. While the first block corresponds to processes of the kind $\pi + \pi \to \rho + \gamma$, the second and third block consist of $\pi + \rho \to \pi + \gamma$  processes. The latter are only different with regard to the mediating particles, ($\pi, \rho, a_1$) mesons in the second and the $\omega$ meson in the third block. Note further that while processes (\ref{C12}) and (\ref{C13}) may be mediated by both, ($\pi, \rho, a_1$) and $\omega$ meson, process (\ref{C11}) only occurs through exchange of ($\pi, \rho, a_1$) mesons and process (\ref{C14}) only though exchange of the $\omega$ meson. ($\pi, \rho, a_1$)-mediated and $\omega$-mediated channels are treated separately to be in accordance with \citep{Turbide:2003si}, where some production channels are included in the imaginary part of the vector meson spectral density.\\
  \linebreak
  Starting from the Lagrangian in Eq.~(\ref{Lagrangian}), it is possible to derive the Feynman rules and matrix elements for each of the processes above, taking into account all contributing Feynman diagrams. The Feynman diagrams are not listed here, the interested reader is referred to the Appendix of \cite{Turbide_PhD}. It is straightforward to determine the differential cross sections, once the matrix elements are known, through:
  \begin{align}
  	\dfrac{\mathrm{d}\sigma}{\mathrm{d}t} = \dfrac{1}{64 \ \pi s \ p^2_{\text{c.m.}}} \ |
  	\mathcal{M}|^2
  	\label{diff_x_sec}
  \end{align}
  The total cross section is finally determined by integration over $t$. Here, $s$ denotes the square of the center-of-mass energy and $p _{\text{c.m.}}$ the center-of-mass momentum of the binary scattering process. $ |\mathcal{M}|^2$ is the matrix element squared. \\
  Eq.~(\ref{diff_x_sec}) is applied to each of the above listed processes. The results obtained for the total cross sections as a function of $\sqrt{s}$ are presented in Fig.~ \ref{Sigmas}, where the upper plot displays processes (\ref{C21}) and (\ref{C22}), the middle one (\ref{C11}) - (\ref{C13}) and the lower one (\ref{C14}) - (\ref{C16}). The vertical line denotes the kinematic threshold in each specific scattering process.
The cross sections of ($\pi, \rho, a_1$)-mediated processes, as depicted in the upper and center panel of Fig.~\ref{Sigmas}, indicate that except for the $\pi^0 + \rho \to \pi + \gamma$ process, all show a decreasing profile with increasing $\sqrt{s}$ and tend to diverge at their respective thresholds. The $\pi^0 + \rho \to \pi + \gamma$ process on the other hand also diverges at the threshold, but slowly increases with rising center-of-mass energies.
  In contrast to ($\pi, \rho, a_1$)-mediated processes, the cross sections of all $\omega$-mediated processes show a similar behavior. Judging from the lower panel in Fig.~\ref{Sigmas}, they are approximately zero in the vicinity of the threshold and increase with rising $\sqrt{s}$. \\
  \newline
  The cross sections for the above listed photon processes are further implemented into a hadronic transport model to set up a framework capable of investigating the production of photons in heavy-ion collisions. To allow for easy usage of these cross sections within other frameworks, the analytic expressions of the depicted cross sections are available on GitHub: \url{https://github.com/smash-transport/phoxtrot}. See Appendix \ref{app:level1} for further details.


\section{\label{ThermalRate}Thermal Photon Rates from Hadronic Transport}

\subsection{\label{Model} Model description}
  The above determined cross sections are implemented in SMASH. The project is open-source and the code is available on GitHub   \citep{SMASH_github}. SMASH is a newly developed hadronic transport approach with vacuum properties. It is designed for the dynamical description of heavy-ion collisions at low and intermediate energies as well as late, dilute, nonequilibrium stages of high-energy heavy-ion collisions. It provides an effective solution of the relativistic Boltzmann equation by modeling the collision integral through formations and decays of hadronic resonances as well as string excitation and fragmentation. The degrees of freedom include all well-established hadrons listed by the PDG \cite{PDG} up to a mass of $M \approx 2.35$ GeV. As SMASH is designed to satisfy detailed balance, it comprises solely binary collisions; multiparticle decays are thus modeled by means of intermediate resonances. The collision finding algorithm is based on the geometric collision criterion. A thorough description of this approach is provided in \cite{Weil:2016zrk} and a systematic comparison to an analytic solution of the Boltzmann equation can be found in \cite{TINDALL2017532}. 
Furthermore, transport coefficients, dileptons and strangeness production have been studied successfully within SMASH \cite{PhysRevC.97.055204, Hammelmann:2018ath, SMASH_dil, Steinberg:2018jvv}. 

\subsubsection{\label{Photons_SMASH} Photon treatment in SMASH}
  Photons are treated perturbatively in SMASH, which is justified by $\alpha_\mathrm{EM} / \alpha_\mathrm{s} \ll 1$.
A photon process occurs, whenever there is a hadronic interaction of two particles that could potentially produce a photon. That applies to all processes where the incoming particles of any hadronic interaction (elastic or inelastic) are equivalent to the incoming particles of one of the photon processes listed in (\ref{C21}) - (\ref{C16}). The produced photons  are directly printed to a separate output, but not further propagated. Instead, the underlying hadronic reaction is performed as if no photon reaction had taken place. Each produced photon is assigned a specific weight $W$, that scales the production probability in terms of cross section ratios:
  \begin{align}
  	W = \frac{\sigma_{\gamma}}{\sigma_{\mathrm{had}}},
  	\label{weighting_factor}
  \end{align}
  with $\sigma_{\mathrm{had}}$ being the cross section of the underlying hadronic interaction and $\sigma_{\gamma}$ the cross section of the performed photon process. \\
  \newline
  Photons are rare probes in heavy-ion collisions. Consequently, they call for extremely high statistics in order to provide useful results. The perturbative treatment is very useful to this end. In addition, so-called \textit{fractional photons} are implemented in SMASH to artificially increase statistics. The realization follows the implementation in UrQMD \cite{UrQMD:2010_photons}, so that instead of producing one photon, $N_\mathrm{frac}$ fractional photons are produced with different kinematic properties. This is achieved by explicitly sampling the final state particles based on the momentum transfer and scattering angle of the binary collision. The exact value of Mandelstam $t$ is randomly sampled from the $t$ distribution provided by the differential cross section, for each fractional photon. Hence, the weighting factor introduced in Eq.~(\ref{weighting_factor}) needs to be reformulated as
  \begin{align}
  	W = \frac{ \frac{\mathrm{d}\sigma_\gamma}{\mathrm{d}t} \ (t_2 - t_1)}
  				  {N_\mathrm{frac} \ \sigma_{\mathrm{had}}},
    \label{weighting_factor_Nfrac}
  \end{align}
  where $\frac{\mathrm{d}\sigma_\gamma}{\mathrm{d}t}$ denotes the differential cross section of the photon process and $t_1$ and $t_2$ are the lower and upper bound of the Mandelstam $t$ variable, respectively. By means of fractional photons, it is possible to cover significantly more photon phase space with one single underlying hadronic interaction.


\subsection{Results}
  The results presented in this section are, unless stated differently, produced in an infinite matter simulation with SMASH-1.5. It is verified that the medium is in thermal and chemical equilibrium. The degrees of freedom considered for the presented simulation are pions, $\rho$ mesons and the photon, which suffice to describe processes (\ref{C21}) - (\ref{C16}). It is worthwhile mentioning that the $\omega$ and $a_1$ mesons need not be included as degrees of freedom in the simulation since, as an artefact of the perturbative treatment, the exchanged particles are never actually formed. Their rest masses and widths do however enter the computation of the cross sections as input parameters. \\
  Note further that for the first part, the width of the $\rho$ meson is set to zero. This originates from the underlying effective field theory at tree level, in which $\rho$ mesons are assumed to be stable particles. Comparisons to theoretical expectations and to the results presented in \citep{Turbide_PhD} are only possible relying on identical assumptions. A detailed discussion about a possible extension to broad $\rho$ mesons is carried out in Sec. \ref{broad}. Except for Sec. \ref{broad} and Appendix \ref{app:Broad_Rho}, $\Gamma_\rho = 0$ is applied in all test cases. Note further that we use a constant elastic cross section of $\sigma = $ 30 mb instead of resonance formations and decays for the collision finding in the presented test cases. This allows for computationally less expensive simulations while guaranteeing sufficiently high statistics, and is justified by the weighting procedure according to Eqs. (\ref{weighting_factor}) and (\ref{weighting_factor_Nfrac}). \\
  In the following, a collection of results are presented for the thermal photon rate as determined from SMASH in a number of different setups. All simulations are carried out with 100 fractional photons, a runtime of 200 fm per event and are averaged over 100.000 events. The cubic box applied to simulate infinite matter has a length of 10 fm and is initialized with pion and $\rho$ meson multiplicities according to the grand-canonical ensemble.

\subsubsection{\label{Rates} Comparison to theoretical expectation}
\begin{figure}
  	\includegraphics[width=0.45\textwidth]{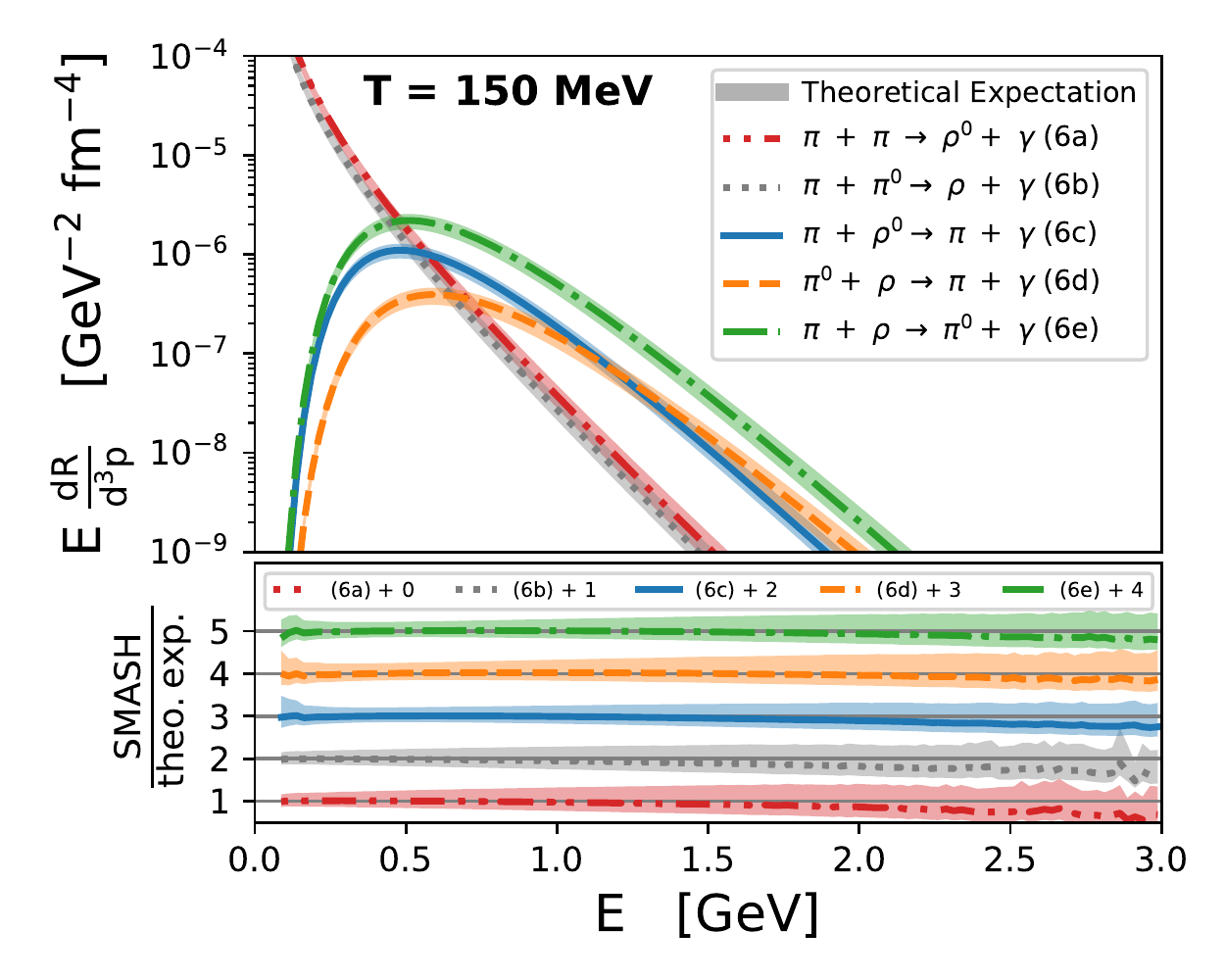}
  	\caption{Comparison of thermal photon rates for ($\pi, \rho,a_1$)-mediated processes, (\ref{C21}) - (\ref{C13}), as
  	determined with SMASH (thin lines) to theoretical expectations (bands) in an infinite matter simulation at a temperature of $T =$ 150 MeV.}
  	\label{Rate_Theory_noOmega}
  \end{figure}

\begin{figure}
  	\includegraphics[width=0.45\textwidth]{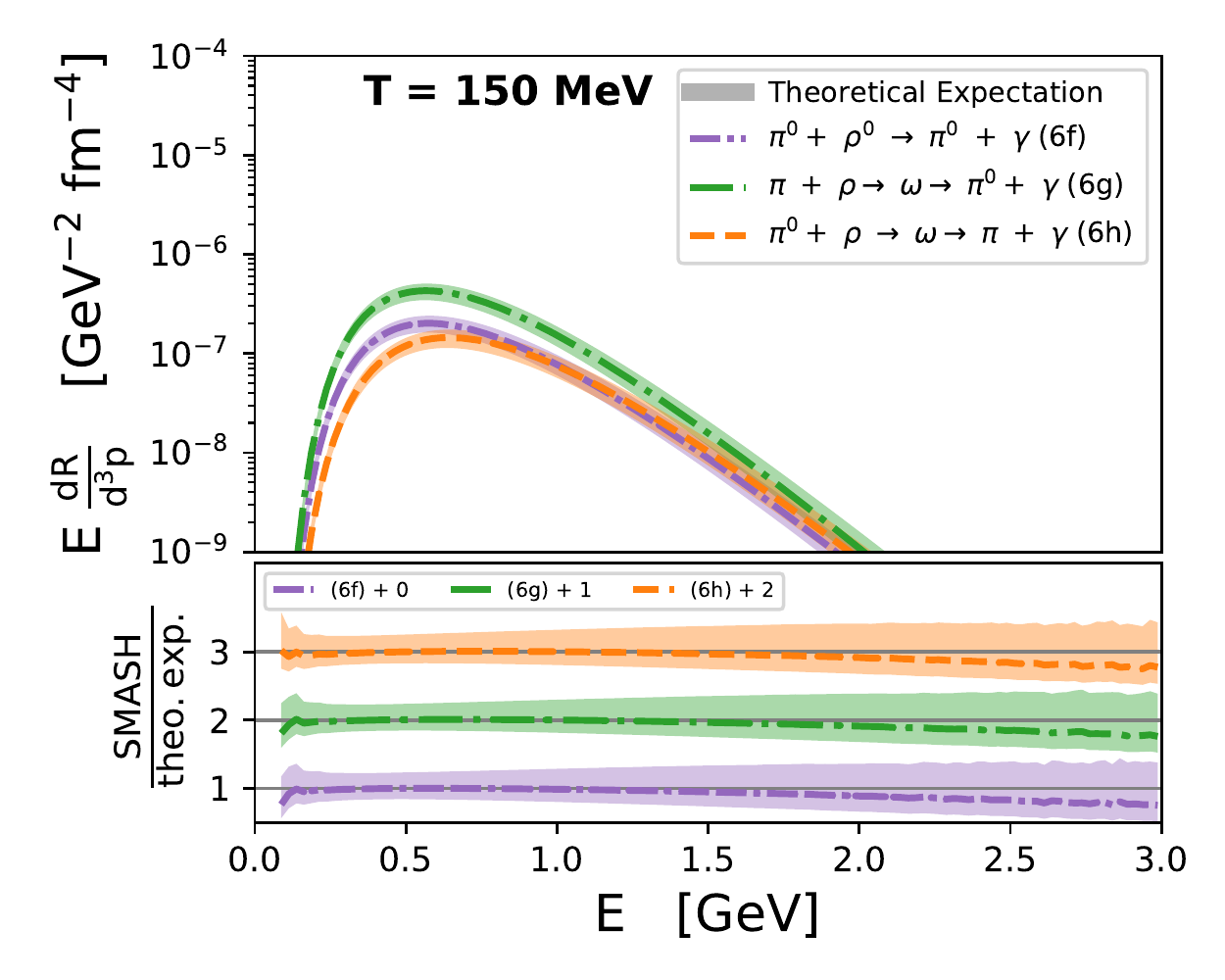}
  	\caption{Comparison of thermal photon rates for $\omega$-mediated processes, (\ref{C14}) - (\ref{C16}), as
  	determined with SMASH (thin lines) to theoretical expectations (bands) in an infinite matter simulation at a temperature of $T =$ 150 MeV. }
  	\label{Rate_Theory_Omega}
  \end{figure}

  First, the functionality of the photon framework in SMASH is verified by means of a comparison to theoretical expectations. The thermal photon rate, i.e. the number of produced photons per unit time and volume, has proven to be most suitable for this comparison. It is well known \citep{Kapusta:1992gv}, that the thermal photon rate for a process of the kind $A + B \to C + \gamma$ is defined as \\
  \begin{align}
  	\begin{split}
  	E \dfrac{dR}{d^3p} = \ & N \ \int (2\pi)^4 \ \delta^4 (p_A ^\mu+ p_B^\mu - p_C^\mu
  	- p^\mu) \ |\mathcal{M}|^2 \\
  	& \times f(E_A) \ f(E_B) \  f(E_C) \ \frac{1}{2(2\pi)^3}\\
  	& \times  \dfrac{d^3p_A}
  	{2E_A(2\pi)^3} \ \dfrac{d^3p_B}{2E_B(2\pi)^3} \ \dfrac{d^3p_C}{2E_C(2\pi)^3},
  	\label{Rate_full}
  	\end{split}
  \end{align} \\
  where $N$ is the degeneracy factor, $f(E_i)$, $E_i$ and $p^\mu_i$ are the distribution functions, the energy and the 4-momenta of particles $A,B,C$ and $p^\mu$ is the photon 4-momentum. \\
  Eq.~(\ref{Rate_full}) is integrated numerically to obtain the theoretical expectations for the photon rates corresponding to processes (\ref{C21}) - (\ref{C16}). Boltzmann equilibrium statistics are applied to match the assumptions of SMASH.
  The theoretically expected rates are depicted as light bands in Fig.~\ref{Rate_Theory_noOmega} and Fig.~\ref{Rate_Theory_Omega}, while the results yielded with SMASH are marked by thinner lines. Note that the effective temperature of the box differs slightly from its initialization temperature, such that the theoretical expectations are computed for the effective temperature. Unfortunately, the  temperature extracted from the momentum distribution in the box is characterized by large uncertainties. In turn, this results in a large error for the theoretically expected photon rate. The bands in the upper plots of Fig.~\ref{Rate_Theory_noOmega} and Fig.~\ref{Rate_Theory_Omega} are therefore drawn between the photon rates corresponding to the lower and upper limit of the effective temperatures. For the ratio in the lower plots of Fig.~\ref{Rate_Theory_noOmega} and Fig.~\ref{Rate_Theory_Omega}, the line  corresponds to the mean of the two aforementioned photon rates, while the error bands again reflect the uncertainty in the temperature determination. The ratios displayed in the lower panel are further scaled linearly to increase readability.\\
The SMASH simulation is carried out at a temperature of $T = 150$ MeV, where strongly interacting matter is expected to be of hadronic origin. As previous works have usually computed photon rates at a temperature of $T = 200$ MeV, SMASH results for this temperature are provided in Appendix \ref{app:Rates200MeV}.
 In Fig.~\ref{Rate_Theory_noOmega}, the thermal photon rate as a function of the photon energy is presented for $(\pi, \rho, a_1)$-mediated processes (\ref{C21})-(\ref{C13}). Within uncertainties, an excellent agreement is found between the results from SMASH and the theoretical expectations. 
In Fig.~\ref{Rate_Theory_Omega}, the thermal photon rate as a function of the photon energy is presented for $\omega$-mediated processes (\ref{C14})-(\ref{C16}). Note that on purpose the scale of the y-axis is identical to the one in Fig.~\ref{Rate_Theory_noOmega} to allow for an easy comparison of the order of magnitude. Again, an excellent agreement is observed within uncertainties. \\
\newline
It can be concluded that the SMASH results presented in Fig.~\ref{Rate_Theory_noOmega} and Fig.~\ref{Rate_Theory_Omega} coincide impressively well with theoretical expectations. Hence, the photon treatment and the underlying dynamics in SMASH are validated. Most importantly however, the nearly perfect agreement verifies the cross section calculations presented above.

\subsubsection{\label{FF}Introduction of form factors}
\begin{figure}
  	\includegraphics[width=0.45\textwidth]{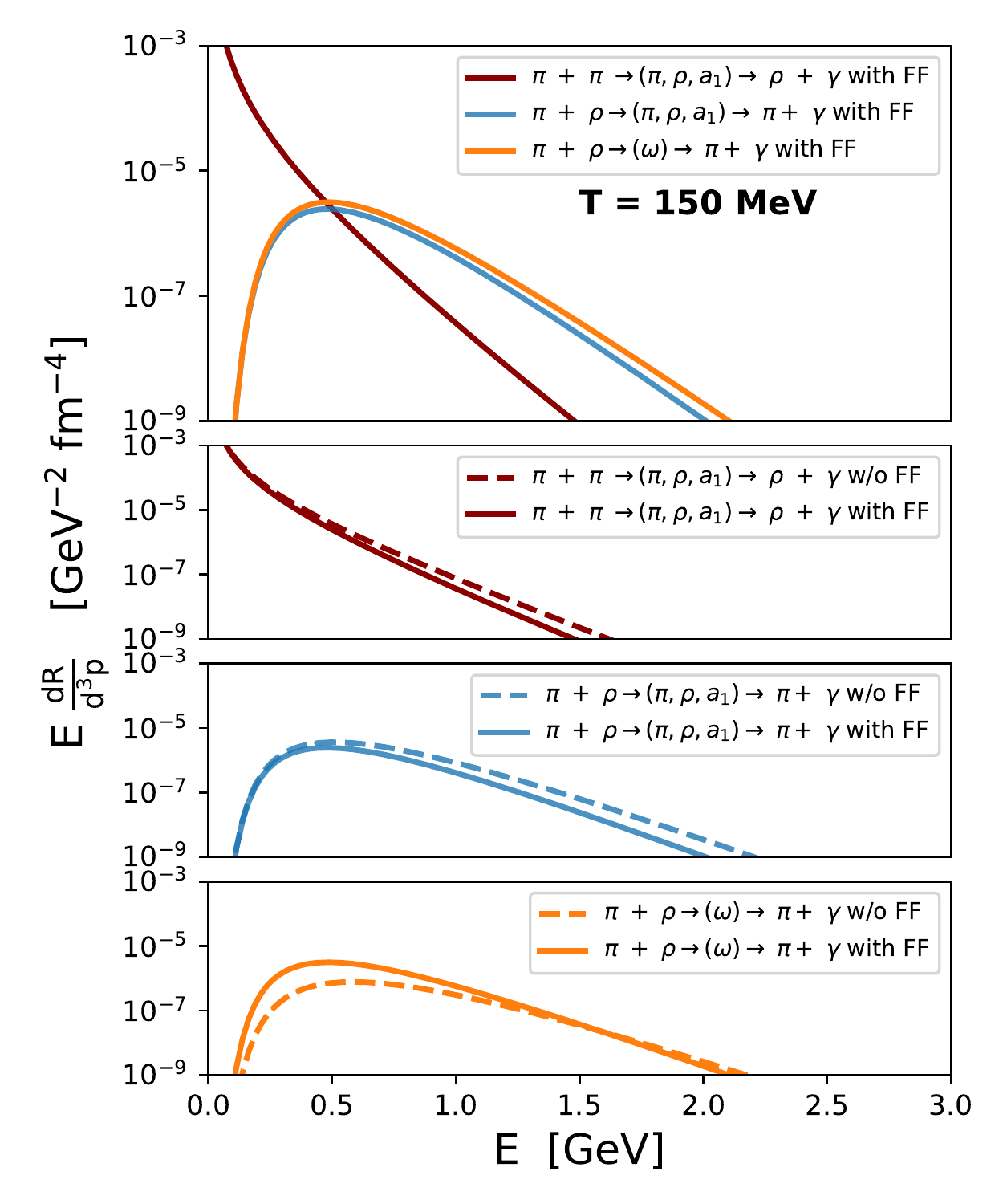}
  	\caption{Thermal photon rates of combined $\pi + \pi \to (\pi, \rho,a_1)  \to \rho + \gamma$ (red), $\pi + \rho \to (\pi, \rho,a_1) \to \pi + \gamma$ (blue) and $\pi + \rho \to \omega \to \pi + \gamma$ (orange) production channels without form factors (dashed lines) and with form factors (solid lines) at $T =$ 150 MeV. The upper plot shows the resulting photon rates with form factors for all three contributions.}
  	\label{Introduction_FF}
  \end{figure}

  The results discussed in the previous section are now improved by properly including form factors following the description in section \ref{Theory}. The results are presented in Fig.~\ref{Introduction_FF}, where the different photon production processes are grouped into 3 categories: Processes (\ref{C21}) and (\ref{C22}) are combined in the red curve and correspond to ($\pi, \rho,a_1$)-mediated $\pi + \pi \to \rho + \gamma$ processes. (\ref{C11}), (\ref{C12}), (\ref{C13}) are combined in the blue curve to account for all ($\pi, \rho,a_1$)-mediated $\pi + \rho \to \pi + \gamma$ processes. The orange curve contains the corresponding $\omega$-mediated processes of this channel, (\ref{C14}), (\ref{C15}) and (\ref{C16}). While the upper plot shows the resulting photon rates with form factors, the three lower plots demonstrate the effect of including form factors, for each of the three groups individually.
  It can be observed that the inclusion of form factors results in a reduction of the photon rate for $\pi + \pi \to \rho + \gamma$ and $\pi + \rho \to (\pi, \rho,a_1) \to \pi + \gamma$ processes, while for $\pi + \rho \to \omega \to \pi + \gamma$ processes, it is reduced only in the high-energy region, but significantly enhanced for low photon energies. Generally, the consideration of form factors causes a decrease of the photon rate which is more pronounced for higher photon energies. For $\omega$ mediated processes however there is an additional ingredient: The coupling constant at the $\pi - \rho - \omega$ vertex is significantly higher once form factors are applied. This is due to the unambiguous relation between the matrix element and the decay width of the $\omega$-meson for the $\omega \to \pi + \gamma$ decay. The latter is known from experiment and requires an adjustment of the coupling constant, from 11.93 GeV$^{-1}$ to 22.6 GeV$^{-1}$. It can be observed, that for photon energies $E \lesssim 1.5$ GeV, the response of the $\omega$-mediated photon rate to the introduction of form factors is completely governed by the increased coupling constant, resulting in a much greater photon production than in the case without form factors. It is only for high photon energies, $E \gtrsim 1.5$ GeV, that the decreasing nature of form factors surpasses the effect of an increased coupling constant. \\
  In fact, the contribution to the total photon rate by $\omega$-mediated processes in $\pi + \rho \to \pi + \gamma$ production channels is found to be on the same order of magnitude as the contribution by ($\pi, \rho,a_1$)-mediated processes, once form factors are taken into consideration. This clearly indicates that photon production channels involving $\omega$ mesons contribute significantly to the total photon production and should therefore not be neglected. Similar conclusions regarding the importance of photon production channels involving $\omega$ mesons, albeit as initial or final state particles, have already been drawn in \citep{Rapp_omega}. \\
  \begin{figure}
  	\includegraphics[width=0.45\textwidth]{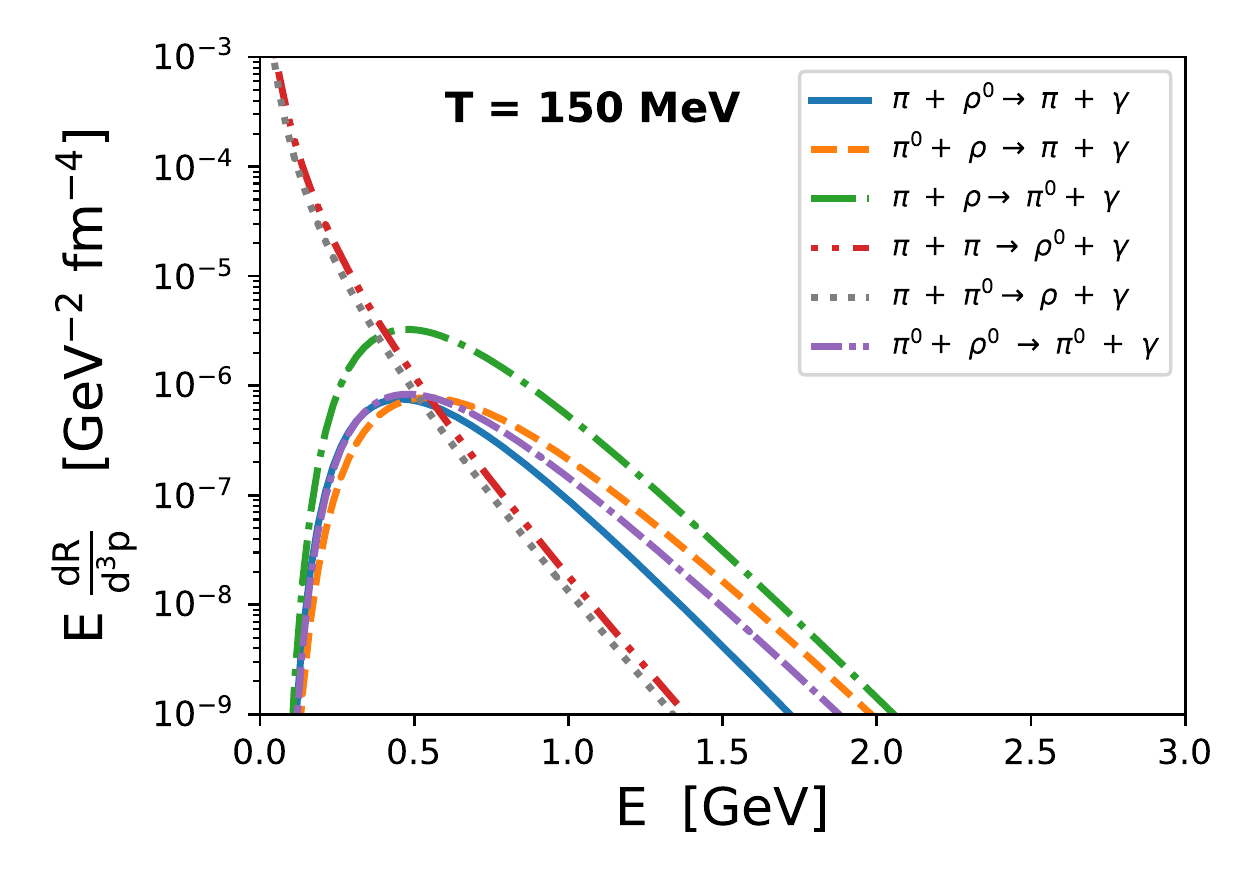}
  	\caption{Thermal photon rates of combined ($\pi, \rho, a_1$)-mediated and $\omega$-mediated processes at a temperature of $T = 150$ MeV from an infinite matter simulation, carried out with SMASH. Form factors are included.}
  	\label{Rates_SUM}
  \end{figure}
  In the following, the cross sections of ($\pi, \rho, a_1$)-mediated and $\omega$-mediated processes with identical initial and final state particles ((\ref{C12}) and (\ref{C16}), (\ref{C13}) and (\ref{C15})) are summed up incoherently, but considering their respective form factors, to define the total cross sections of these production channels. The resulting photon rates are displayed in Fig.~\ref{Rates_SUM}, where the low-energy region is still dominated by $\pi + \pi \to \rho + \gamma$ processes and the high-energy region by $\pi  + \rho \to \pi + \gamma$ processes. \\
 \begin{figure}
  	\includegraphics[width=0.45\textwidth]{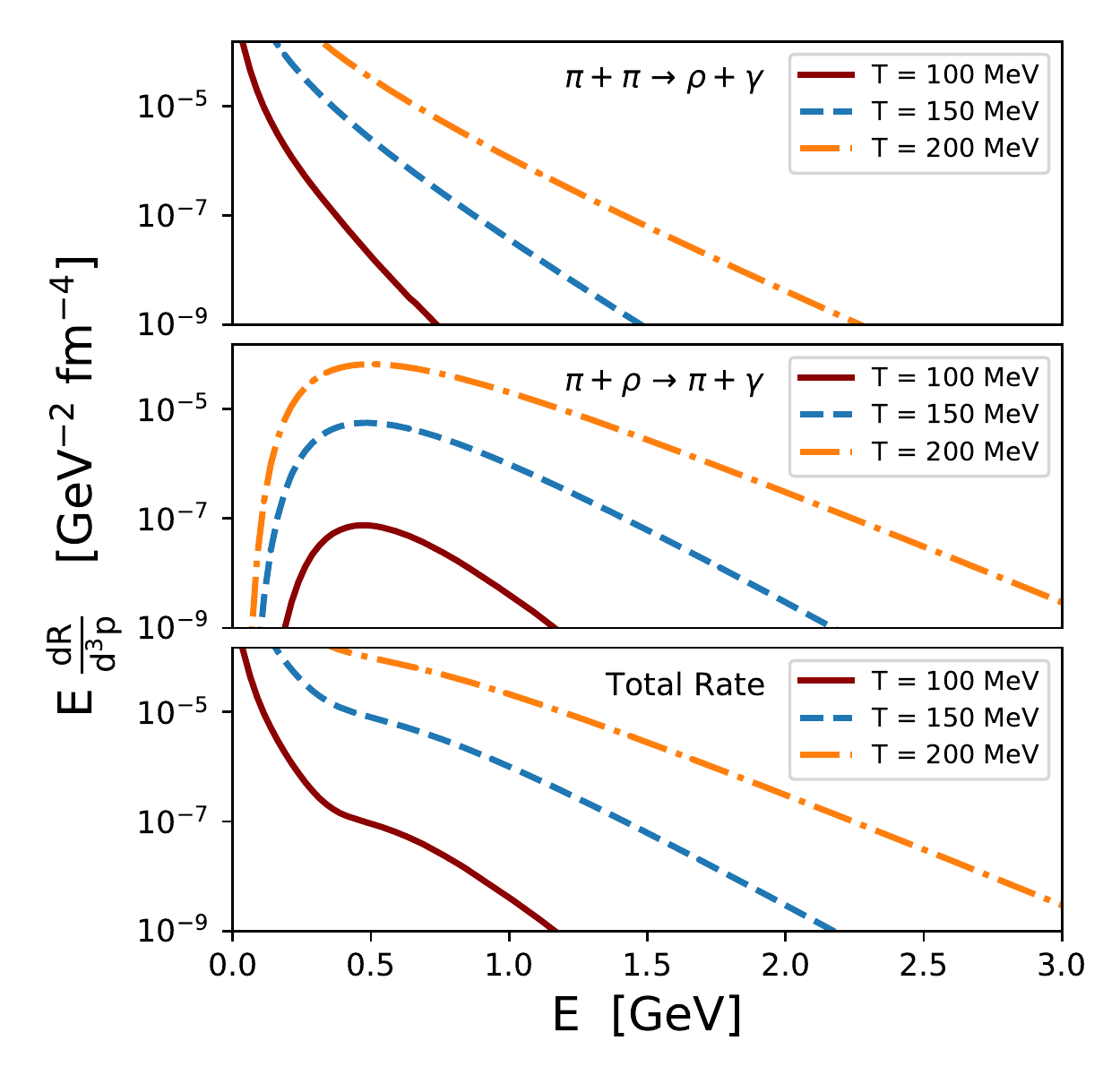}
  	\caption{Temperature scaling of the thermal photon rate for $\pi + \pi \to \rho + \gamma$ processes (upper panel), for $\pi + \rho \to \pi + \gamma$ processes (center panel), and the total contribution (lower panel). Form factors are included. The y-axis is cut since the rates are growing rapidly at small energies.}
  	\label{Temp_Scaling}
  \end{figure}
It is further instructive to investigate the dependence of the photon rate on the temperature of the system. Such a comparison is performed in Fig.~\ref{Temp_Scaling}, where the photon rate in $\pi + \pi \ \to \ \rho + \gamma$ processes (upper),  $\pi + \rho \ \to \ \pi + \gamma$ processes (center) and the total contribution (lower) is displayed for three different temperatures: $T = 100$ MeV (full, red), $T = 150$ MeV (dashed, blue) and $T = 200$ MeV (dot-dashed, orange). It can be observed that the photon rate increases with rising temperatures, by approximately three orders of magnitude between $T = $ 100 MeV and $T =$ 200 MeV, independently of the scattering process. \\

\subsubsection{\label{Turbide_Comp}Comparison to parametrized photon rates}
\begin{figure}
  	\includegraphics[width=0.45\textwidth]{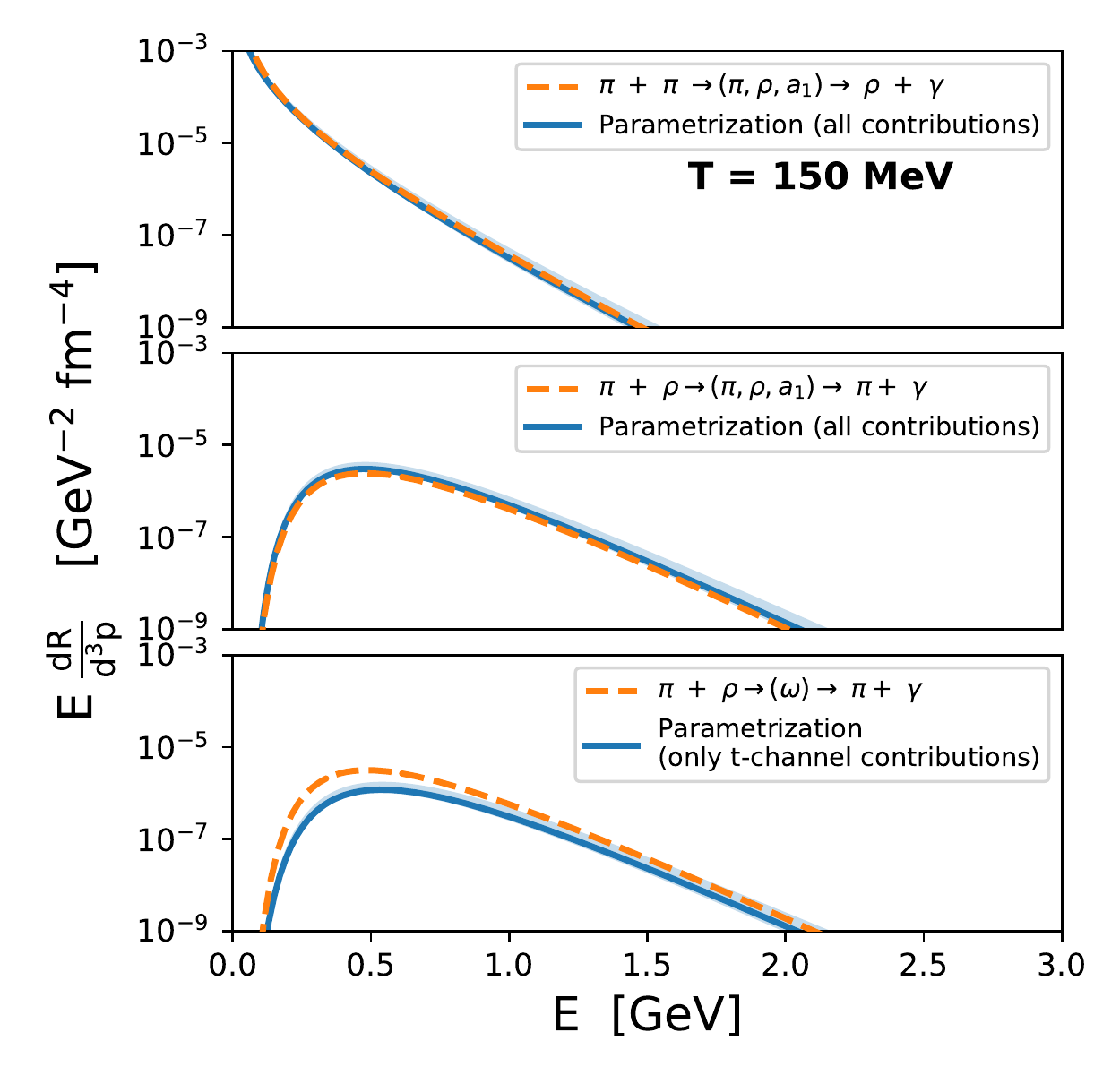}
  	\caption{Comparison of thermal photon rates within SMASH (dashed lines) to parametrizations of these rates (solid lines), taken from \citep{Turbide_PhD}, at $T =$ 150 MeV.}
  	\label{Comp_Turbide_Param}
  \end{figure}

The authors of \citep{Turbide_PhD, Turbide:2003si} have further provided parametrizations of the thermal photon rates corresponding to the framework and cross sections implemented in SMASH. These rates are for example applied in \mbox{MUSIC}, a 3 + 1D hydrodynamic calculation for heavy-ion collisions \citep{MUSIC_base, MUSIC_bulkVisc}, to describe photon production in the hadronic phase. Form factors are included in their calculations. In Fig.~\ref{Comp_Turbide_Param}, the provided parametrizations are compared to the thermal photon rates obtained within SMASH at a temperature of $T =$ 150 MeV. The parametrizations are displayed with solid lines, the SMASH results with dashed lines. They are grouped into $\pi + \pi \to \rho + \gamma$ processes (upper), $\pi + \rho \to (\pi,\rho,a_1) \to \pi + \gamma$ processes (center) and $\pi + \rho \to \omega \to \pi + \gamma$ processes (lower). As discussed in Sec. \ref{Rates}, the effective temperature of the system is characterized by large uncertainties. Consequently, the bands around the parametrized rates represent the range of the parametrizations corresponding to the lower and upper limit of the extracted temperature.
 It can be observed that while there is a very good agreement for the $\pi + \pi \to \rho + \gamma$ and $\pi + \rho \to (\pi,\rho,a_1) \to \pi + \gamma$ channels, there is a significant discrepancy in the case of the $\pi + \rho \to \omega \to \pi + \gamma$ channel. This is due to different assumptions underlying the SMASH calculation and the parametrizations. While the above derived cross sections for $\omega$-mediated processes account for all contributing Feynman diagrams (s-, and t-channel), the parametrizations only contain the t-channels. The s-channels are absorbed in the in-medium $\rho$ spectral function \citep{Turbide:2003si}, while SMASH relies on vacuum properties. An overestimation with SMASH is therefore expected.

\subsubsection{\label{Kapusta_Sec} Comparison to other sources of direct photons}
\begin{figure}
  	\includegraphics[width=0.45\textwidth]{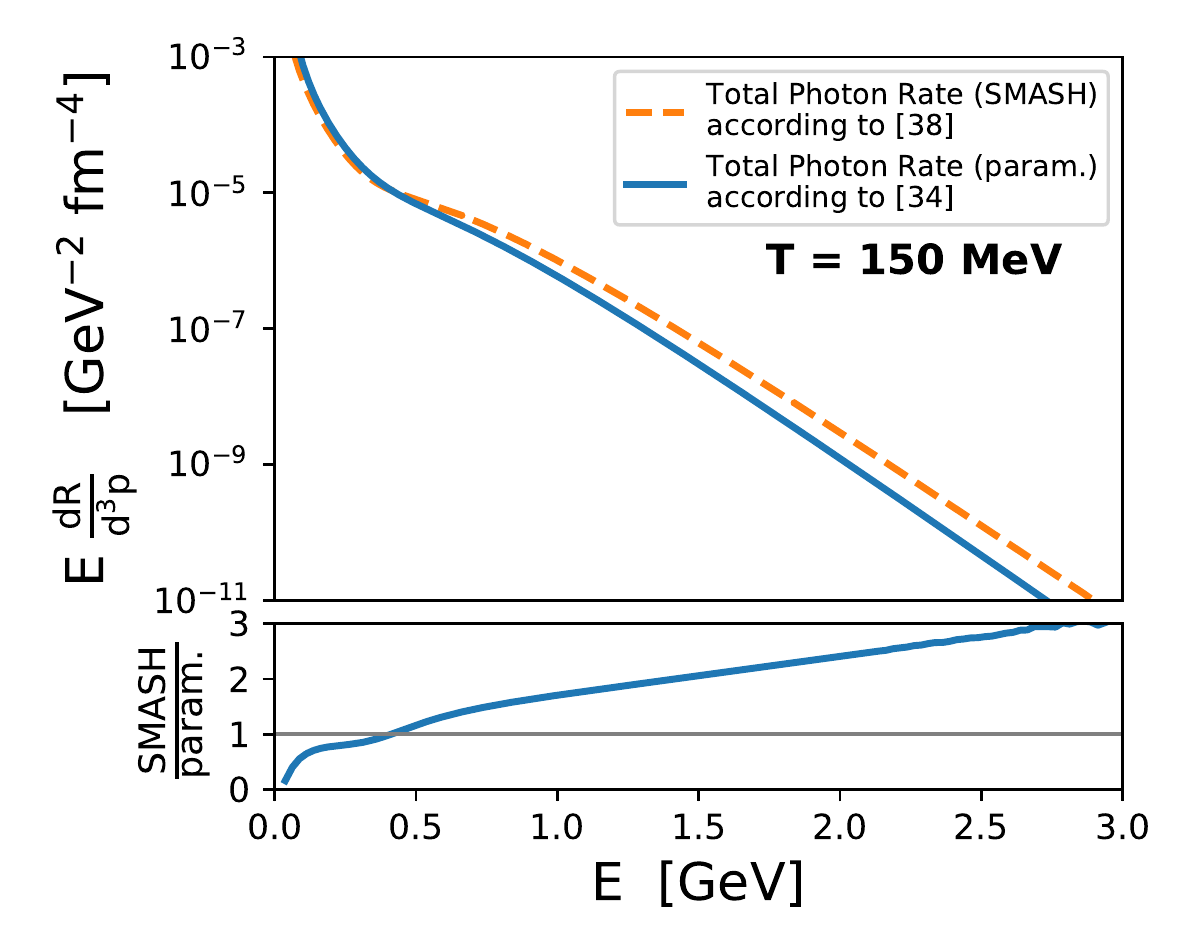}
  	\caption{Comparison of thermal photon rates from SMASH following the logic in \citep{Turbide_PhD,Turbide:2003si} (dashed line) to parametrization of the photon rates used in other transport models, provided in \citep{Kapusta:1992gv} (solid line). The parametrizations are taken from \cite{Nadeau:1992cn}.}
  		\label{Kapusta_Comp}
  \end{figure}

It is further possible to assess the difference between the photon rates from SMASH based on \citep{Turbide_PhD,Turbide:2003si}, and the photon rates in \citep{Kapusta:1992gv} that are applied in other hadronic transport approaches \citep{PHSD:2016_photons, UrQMD:2010_photons}. Note though that both sets of photon rates are determined within different effective field theories relying on different degrees of freedom. As a consequence, the scattering processes in SMASH are mediated by either $\pi, \ \rho, \ a_1$ or $\omega$ mesons while the photon rates in \citep{Kapusta:1992gv} can be mediated through exchange of $\pi$ or $\rho$ mesons only. The parametrizations of the photon rates in \citep{Kapusta:1992gv} are provided in \citep{Nadeau:1992cn}, but are modified by the form factors defined in Eqs.~(\ref{hadronic_dipole_FF}),(\ref{FF_pi}),(\ref{FF_omega}) to allow for an appropriate comparison. In Fig.~\ref{Kapusta_Comp} they are summed up (solid line) and compared to the total photon rate from SMASH (dashed line). As expected, the photon rates resulting from the different frameworks are not identical. The ratio in the lower plot shows that the framework described in \citep{Kapusta:1992gv} provides higher photon rates than SMASH for $E \lesssim 0.4$ GeV, and vice versa; where the discrepancy is more pronounced  for rising photon energies. These comparisons show the importance of the $a_1$- and $\omega$-mediated processes, which are not included in the photon rates from \citep{Kapusta:1992gv}. Since the two rates differ considerably, differences in the photon production in the late hadronic stage of heavy-ion collisions are also expected.  \\
Another well-established set of photon rates was derived by Arnold, Moore and Yaffe in \citep{AMY_Rates:2001, AMY_Rates:2001_2}. In contrast to the photon rates in \citep{Turbide:2003si} and \citep{Kapusta:1992gv}, the AMY photon rates describe photon production in a partonic instead of a hadronic medium. Previously, a coincidence of these rates has been observed above the critical temperature \citep{Gale:2014_Photons_Dileptons_QGP+HM}, whereas a priori one might not expect either approach to be viable close to $T_\mathrm{c}$. 
  \begin{figure}
  	\includegraphics[width=0.45\textwidth]{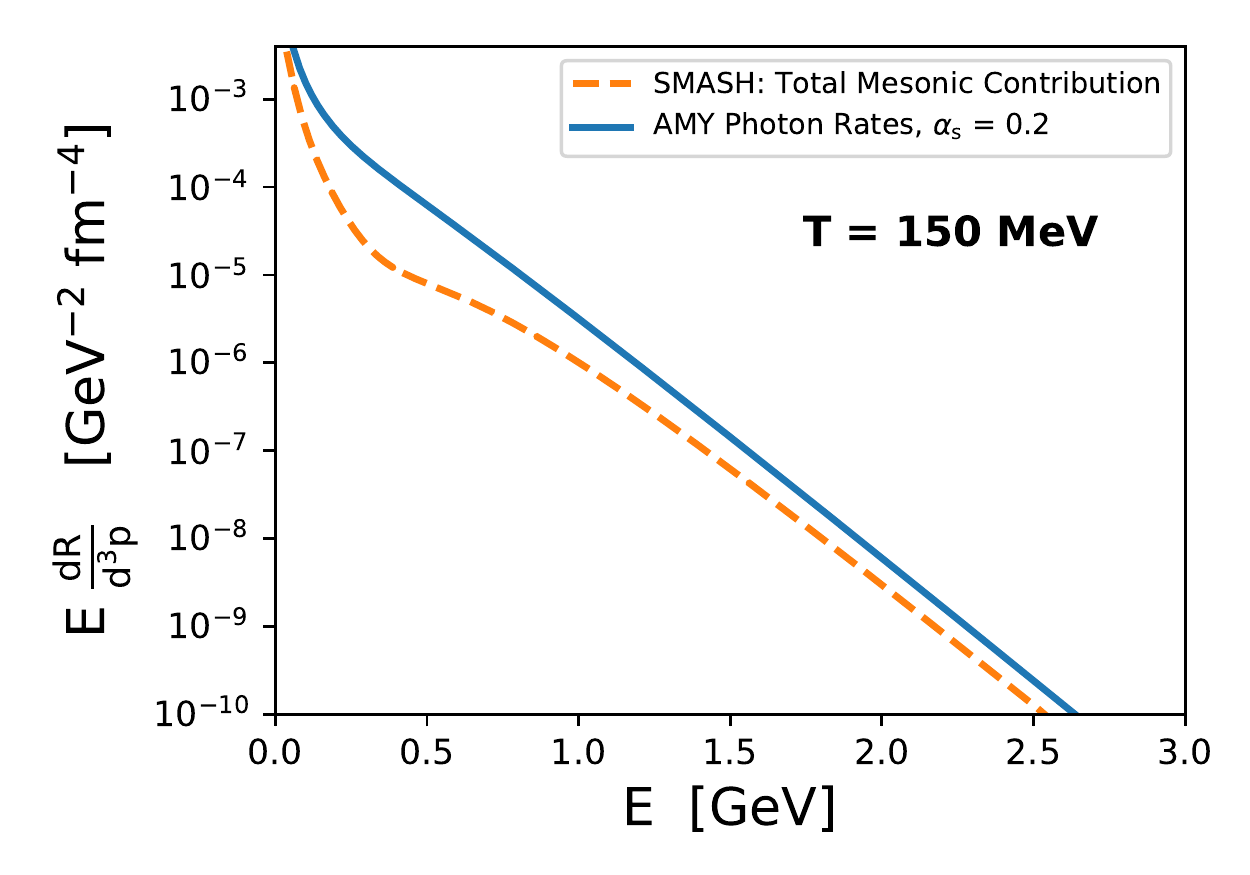}
  	\caption{Comparison of thermal hadronic photon rates from SMASH (dashed) to partonic photon rates (solid) described within the AMY framework \citep{AMY_Rates:2001} at a temperature of $T =$ 150 MeV. For the AMY parametrizations, $\alpha_\mathrm{s}$ = 0.2 and 2-flavor QCD are applied.}
  	  		\label{AMY_Comp}
  \end{figure} 
In Fig. \ref{AMY_Comp}, a comparison between the thermal hadronic photon rate from SMASH (dashed) and the partonic AMY photon rate (solid) is performed at a temperature of $T = 150$ MeV. For the AMY rates, $\alpha_\mathrm{s}$ = 0.2 and $N_\mathrm{flavor}$ = 2 are assumed. Note though, that the applied value for $\alpha_\mathrm{s}$ is merely a rough estimate, but sufficient to draw a qualitative comparison. It can be concluded that while both rates decrease with rising photon energies, the AMY contributions are significantly higher than the mesonic contributions from SMASH. A considerable deficit of the SMASH photon rate with respect to the AMY rates can further be observed for $E \approx$ 0.5 GeV. It is however important to mention that the performed comparison is characterized by differences between the two frameworks and thus allows for a very qualitative comparison only. While the AMY parametrizations account for scattering processes, annihilation processes, Bremsstrahlung processes and the Landau-Pomeranchuk-Migdal (LPM) effect \citep{LPM-effect:Aurenche2000}, SMASH is based purely on binary scatterings; lacking the LPM effect as well as annihilation processes and Bremsstrahlung. Including the latter is important but left for future work. Keeping these caveats in mind, it is interesting to find both frameworks providing photon rates at similar orders of magnitude.

\subsubsection{\label{broad} Extension to broad $\rho$ meson}
  The so far presented calculations rely on a stable $\rho$ meson, mainly to be in accordance with the underlying field theory at tree level. Experimental measurements have however shown that $\rho$ mesons are characterized by a significant finite width of $\Gamma_\rho = 0.149$ GeV \cite{PDG}. A more realistic description within theoretical calculations therefore calls for a consideration of this nonvanishing width in the computation. While such considerations are challenging within the underlying field theory, it is rather easily feasible within SMASH. As such, the same infinite matter simulation as previously presented is performed with $\Gamma_\rho = 0.149$ GeV instead of $\Gamma_\rho = 0$ GeV. The mass distribution of $\rho$ mesons then follows a relativistic Breit-Wigner distribution:
\begin{align}
\mathcal{A}(m) = \frac{2\mathcal{N}}{\pi} \frac{m^2 \Gamma(m)}{(m^2 - M_0^2)^2 + m^2 \Gamma(m)^2} ,
\end{align}
 with $\mathcal{N}$ being the degeneracy factor, $m$ the actual off-shell mass, $\Gamma (m)$ the mass-dependent width and $M_0$ the pole mass. Further information about the resonance treatment in SMASH is provided in \citep{Weil:2016zrk}. There is however one caveat to simply applying this photon framework to broad $\rho$ mesons: The photon production processes (\ref{C21}), (\ref{C12}) and (\ref{C13}) are each characterized by one contributing diagram in which the scattering process $\pi + \rho \to \pi + \gamma$ is also mediated by a $\rho$ meson. This means, there are two $\rho$ mesons participating in the scattering, one in the initial state, the other in the intermediate state. Those $\rho$ mesons could in principle have different masses, which entails problems in the conservation of the electromagnetic current $J^\mu$, such that 
\begin{align}
\partial_\mu J^\mu \neq 0
\end{align}
This is related to some contributions in the matrix elements being proportional to 
  \begin{align}
  \Delta \ \equiv \ \frac{m_\rho^2 - u}{M_\rho^2 - u},
  \label{Delta}
  \end{align}
  where $m_\rho$ denotes the mass of the incoming $\rho$ meson and $M_\rho$ the mass of the intermediate one. Current conservation is only assured for $\Delta = 1$. In the case of $\Gamma_\rho = 0$ GeV, it applies that
  \begin{align}
  m_\rho = M_\rho \qquad      \Leftrightarrow  \qquad \Delta = 1
  \end{align}
  whereas in the case of broad $\rho$ mesons,
    \begin{align}
  m_\rho \neq M_\rho \qquad      \Leftrightarrow  \qquad \Delta \neq 1.
  \label{Gauge_breaking_delta}
  \end{align}
 generally holds. Eq.~(\ref{Gauge_breaking_delta}) is particularly problematic in view of current conservation, considering that $\Delta \neq 1$ implies the electromagnetic current is not conserved in processes (\ref{C21}), (\ref{C12}) and (\ref{C13}). To circumvent this problem, the cross sections used in the photon producing scattering processes, are computed with $m_\rho = M_\rho$, such that $\Delta = 1$ is enforced, and current conservation is assured. It is obvious that such a treatment is physics-wise not entirely complete; the incoming and the intermediate $\rho$ meson can in principle have different masses. At the same time, their masses are on average expected to be the pole masses, which suggests the average difference between $m_\rho$ and $M_\rho$ can be assumed to be small and $\Delta$ approximated with unity.
This justifies the assumption $m_\rho = M_\rho$ whilst at the same time giving rise to a systematic error of extending the presented approach to broad $\rho$ mesons. Further discussion of this issue can be found in Appendix \ref{app:Broad_Rho}.  \\
  \begin{figure}
  	\includegraphics[width=0.45\textwidth]{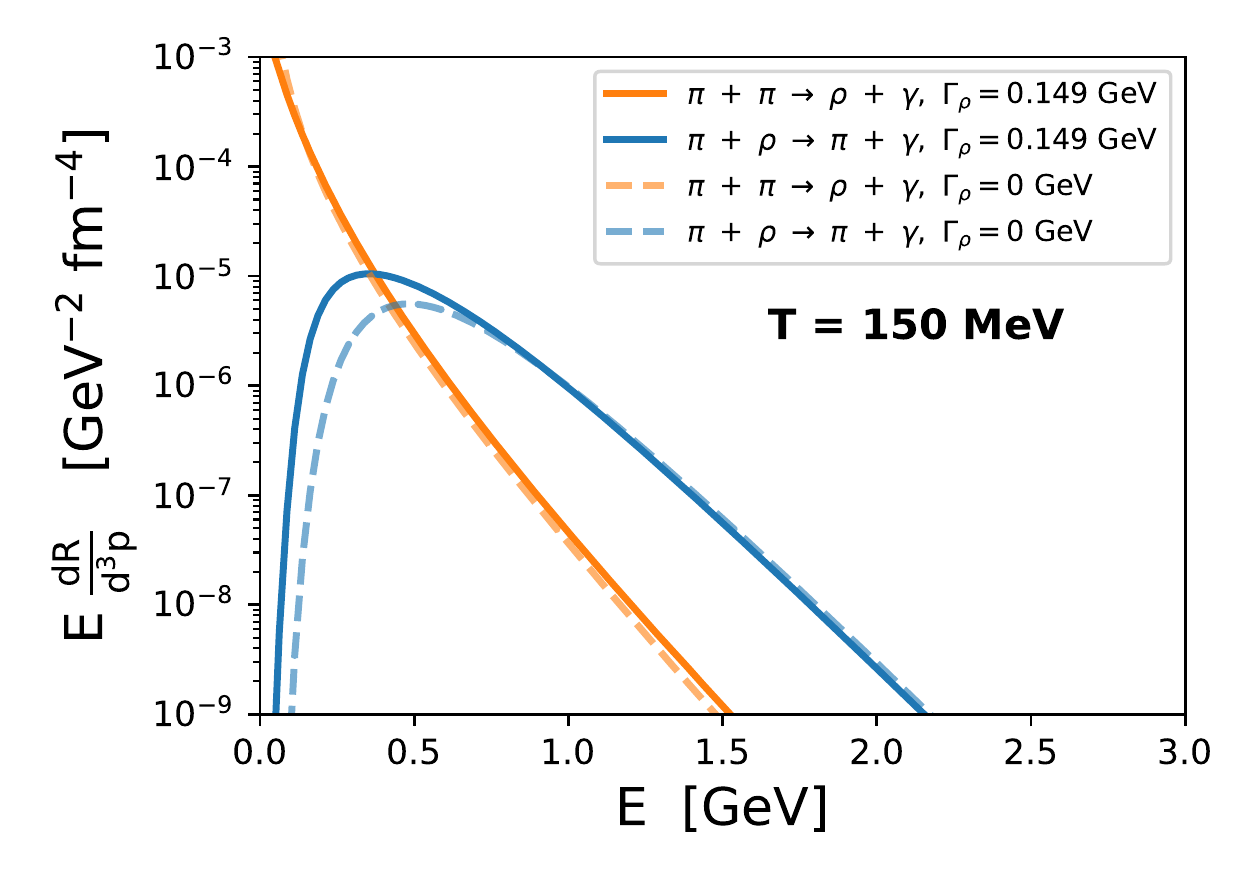}
  	\caption{Effect of considering finite-width $\rho$ mesons on the thermal photon rate in $\pi + \pi \to \rho + \gamma$ processes (red) 
  				and $\pi + \rho \to \pi + \gamma$ processes (blue). Computations with stable $\rho$ mesons are displayed with solid lines, 
  				those with finite $\rho$ mesons with dashed lines. }
  	\label{Broad_Photon_Rates}			
  \end{figure}
 Fig.~\ref{Broad_Photon_Rates} shows the thermal photon rate for combined $\pi + \pi \ \to \ \rho + \gamma$ (orange) and $\pi + \rho \ \to \ \pi + \gamma$ (blue) processes with stable (dashed) and broad (solid) $\rho$ mesons. Most notably, a significant enhancement of the photon rate in the region of low photon energies, $E \leq 0.7$ GeV, is observed for $\pi + \rho \ \to \ \pi + \gamma$ processes. At the same time, the photon rate is slightly reduced for higher photon energies. For $\pi + \pi \to \rho + \gamma$ channels, an opposite trend can be observed, albeit significantly less pronounced than in the previous case.\\


\section{\label{Outlook}Conclusion and Outlook}
Cross sections for the production of photons in hadronic processes have been derived from an effective chiral field theory with mesonic degrees of freedom for 8 different production processes. These cross sections are incorporated into a hadronic transport approach (SMASH), useful to simulate low-energy heavy-ion collisions as well as the late and dilute stages of high-energy heavy-ion collisions. The functionality of the presented framework has been verified by an excellent agreement between the thermal photon rates extracted from SMASH and their theoretical expectations. The introduction of form factors has further demonstrated the importance of $\omega$-involving contributions to the thermal photon rate, which supports previous statements made in \citep{Rapp_omega}. It has also been shown that the determined photon rates show a good agreement with parametrizations of the very same rates, as they are being used in hydrodynamic simulations. At the same time, the above presented framework provides slightly higher photon rates than previous works in \citep{Kapusta:1992gv}, which is attributed to the differences in the underlying effective field theories. Finally, the presented framework has been extended to allow for a description of broad instead of stable $\rho$ mesons. A significant enhancement of the thermal photon rate, especially in the region of low photon energies, has been observed. \\
Mesonic photon production is implemented in SMASH and provides good results. Yet, additional contributions to the total photon production need also be considered. Among these are Bremsstrahlung processes, baryonic scattering processes and possibly an extension to the $\omega$-involving processes presented in \citep{Rapp_omega}. Photon production cross sections from processes involving kaons instead of pions can be calculated in an analogous fashion in the future, even though their contribution to the rate is expected to be subleading. In continuation, the photon framework in SMASH can be applied within hybrid models for the description of the late and dilute stages of heavy-ion collisions at RHIC/LHC energies and contribute to the understanding of the 'photon flow puzzle' \citep{Gale:EM_Radiation_Progress_Puzzles}.

\begin{acknowledgements}
  The authors thank J.-F. Paquet, H. van Hees, M. Greif and J. Staudenmaier for fruitful discussions and C. Shen for providing matrix elements. This work was made possible thanks to funding from the Helmholtz Young Investigator Group VH-NG-822 from the Helmholtz Association and GSI, and supported by the Helmholtz International Center for the Facility for Antiproton and Ion Research (HIC for FAIR) within the framework of the Landes-Offensive zur Entwicklung Wissenschaftlich-Oekonomischer Exzellenz (LOEWE) program from the State of Hesse. This project was further supported by the DAAD funded by BMBF with Project-ID 57314610. A.S. acknowledges support by the Stiftung Polytechnische Gesellschaft Frankfurt am Main. J.M.T.-R. was supported by the U.S. Department of Energy under Contract No. DE-FG-88ER40388. C. G. is supported in part by the Natural Sciences and Engineering Research Council of Canada. H.E. acknowledges support by the Deutsche Forschungsgemeinschaft (DFG) through the Grant No. CRC-TR 211 “Strong-interaction matter under extreme conditions”. Computational resources have been provided by the Center for Scientific Computing (CSC) at the Goethe- University of Frankfurt and the GreenCube at GSI. \\
\end{acknowledgements}


\appendix

\section{\label{app:level1}Photon Cross Sections}
The photon cross sections derived within this work are publicly available in C++ -readable format on GitHub. The  PHOXTROT project can be accessed through \url{https://github.com/smash-transport/phoxtrot}. PHOXTROT also provides a framework to easily produce cross section plots by means of cmake. As such, Fig.~\ref{Sigmas}, Fig.~\ref{DiffSigma_PiPi} and Fig.~\ref{DiffSigma_PiRho} were created with of PHOXTROT-1.0.

\section{\label{app:diffTheta}Differential Cross Sections}
\begin{figure}[h]
  \includegraphics[width=0.42\textwidth]{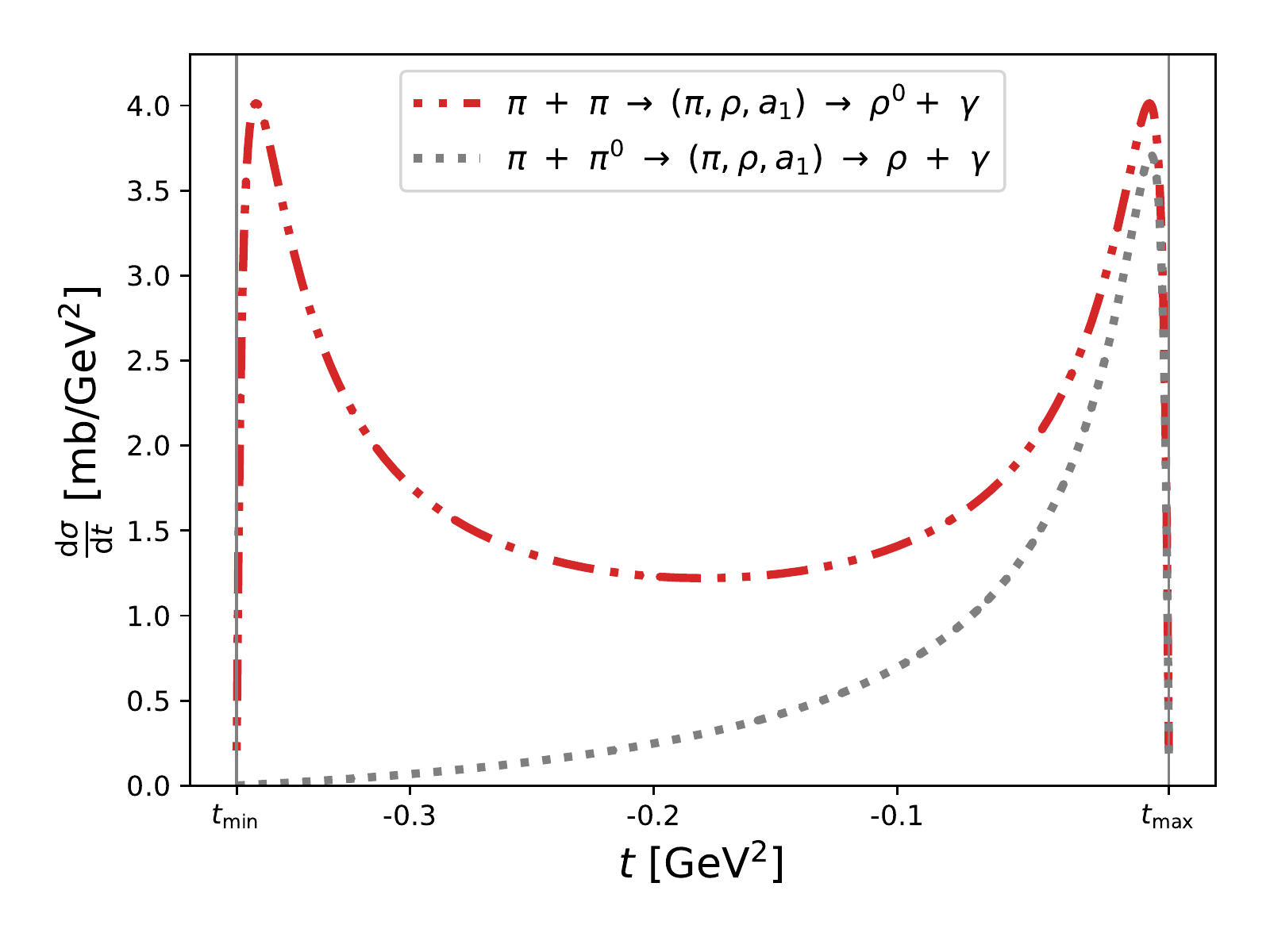} 
  \includegraphics[width=0.42\textwidth]{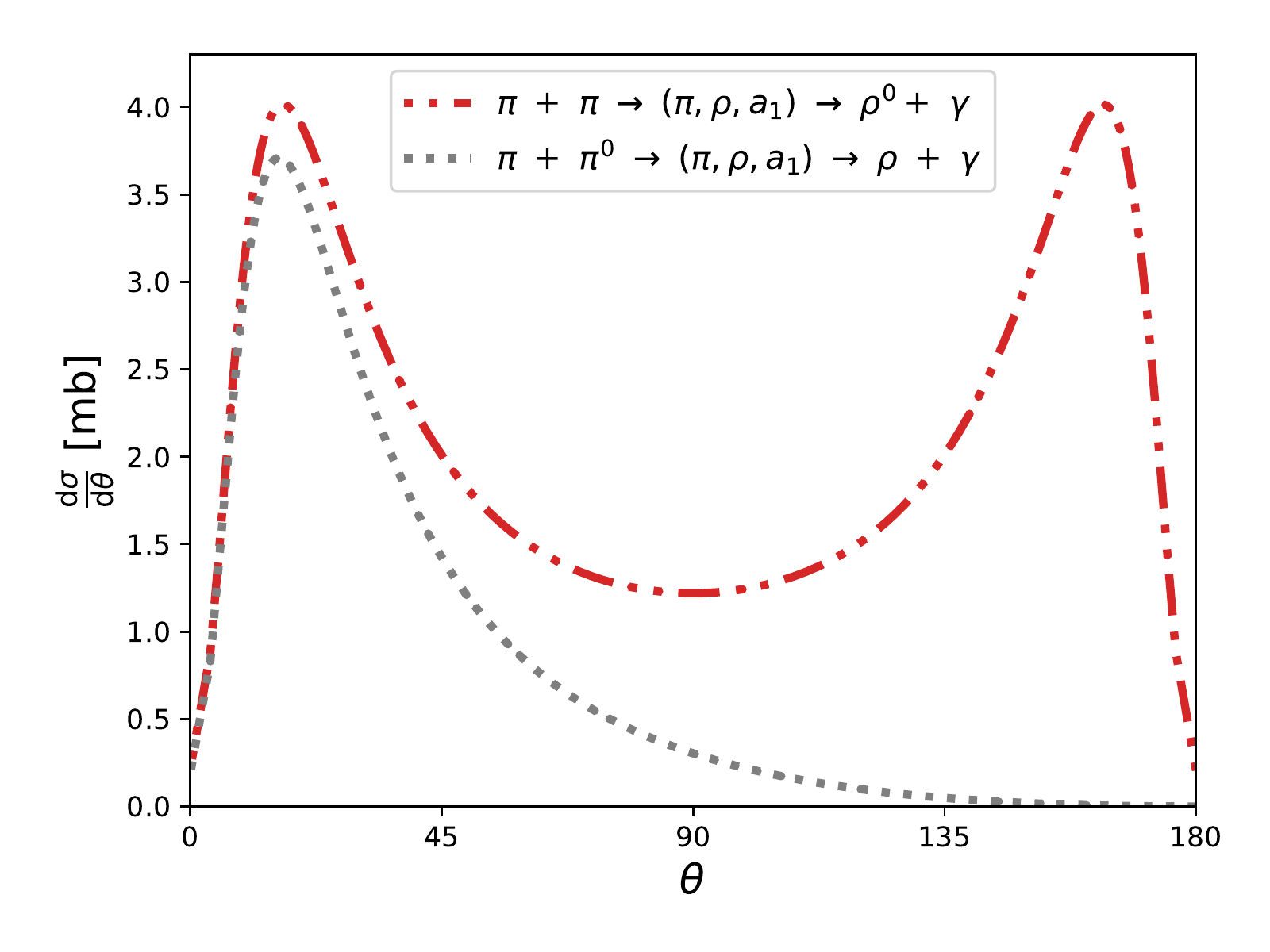}
  	\caption{Differential cross section of $\pi + \pi  \rightarrow \rho + \gamma$ processes as a function of Mandelstam $t$ (upper) and the scattering angle $\theta$ (lower).}
  	\label{DiffSigma_PiPi}
 \end{figure}

\begin{figure}
  	\includegraphics[width=0.42\textwidth]{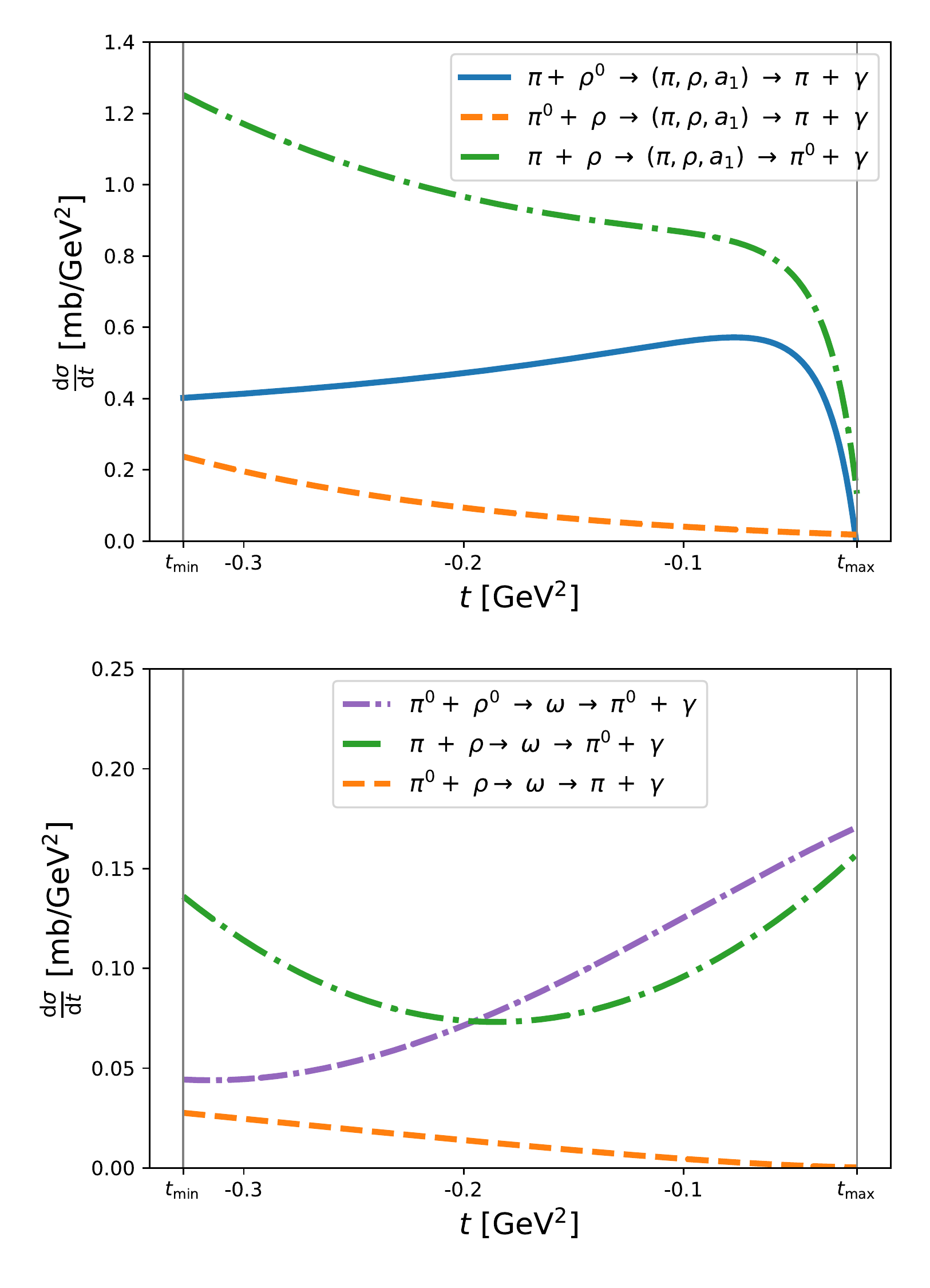} \\[+0.5cm]
  	\includegraphics[width=0.42\textwidth]{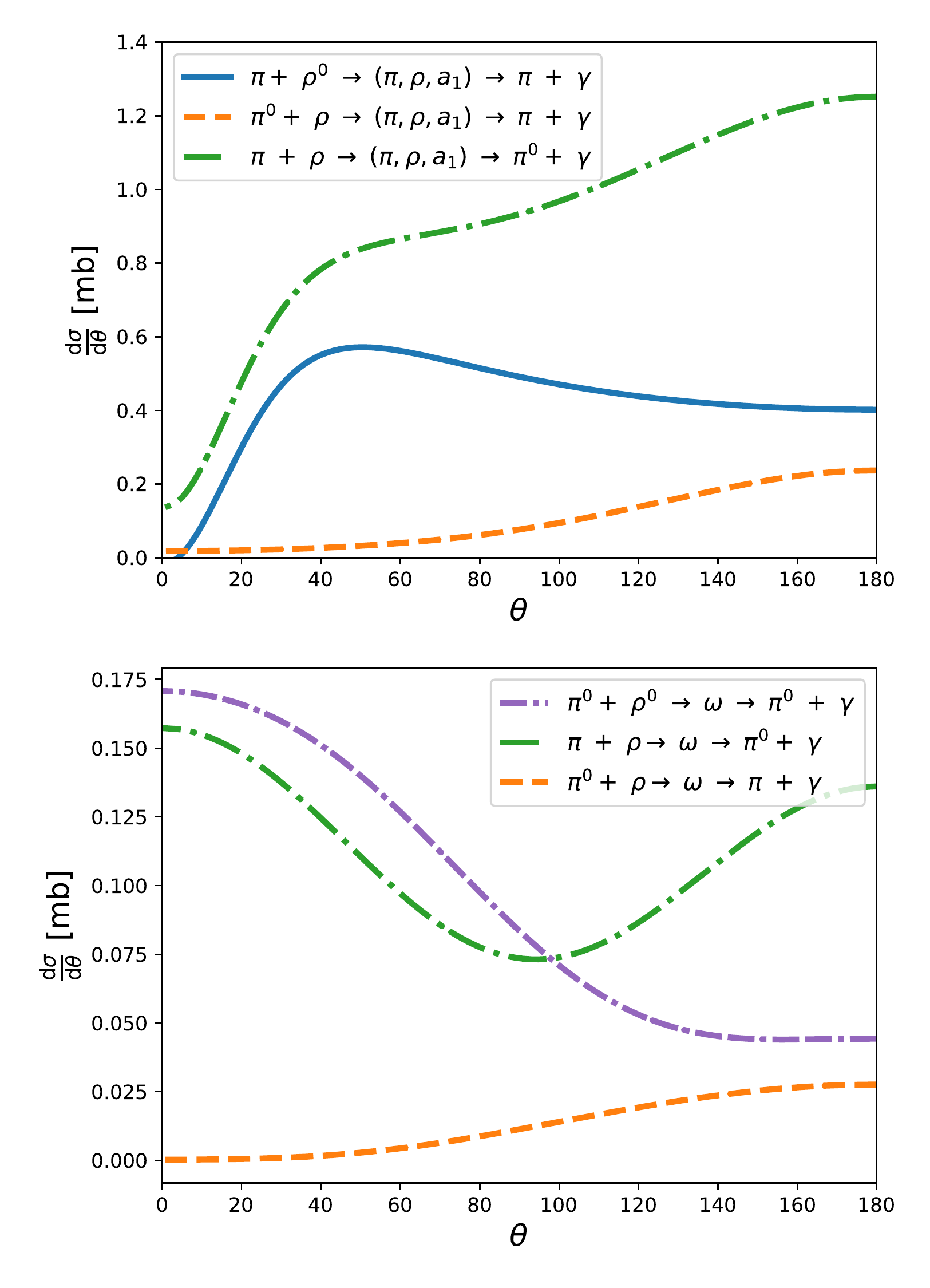}
  \caption{Differential cross sections of ($\pi, \rho, a_1$)-mediated and ($\omega$)-mediated $\pi + \rho \rightarrow \pi + \gamma$ processes as a function of Mandelstam $t$ (upper plots) and of the scattering angle $\theta$ (lower plots).}
  	\label{DiffSigma_PiRho}
  \end{figure}

In addition to the above presented total cross sections, it is further possible to derive the differential cross sections as functions of Mandelstam $t$ or the scattering angle $\theta$ for channels (\ref{C21}) - (\ref{C16}).\\
For completeness, the differential cross sections for \linebreak $\pi + \pi \to \rho + \gamma$ processes are displayed in Fig.~\ref{DiffSigma_PiPi} while those of $\pi + \rho \to (\pi, \rho, a_1 ) \to \pi + \gamma$ processes can be found in Fig.~\ref{DiffSigma_PiRho}. The upper plot in Figure \ref{DiffSigma_PiPi} and the two upper plots in Fig.~\ref{DiffSigma_PiRho} show them as a function of Mandelstam $t$ and the lower ones as a function of the scattering angle $\theta$.  \\

\section{\label{app:Rates200MeV} Photon Rates at $\mathbf{T =}$ 200 MeV}
\begin{figure}
  	\includegraphics[width=0.45\textwidth]{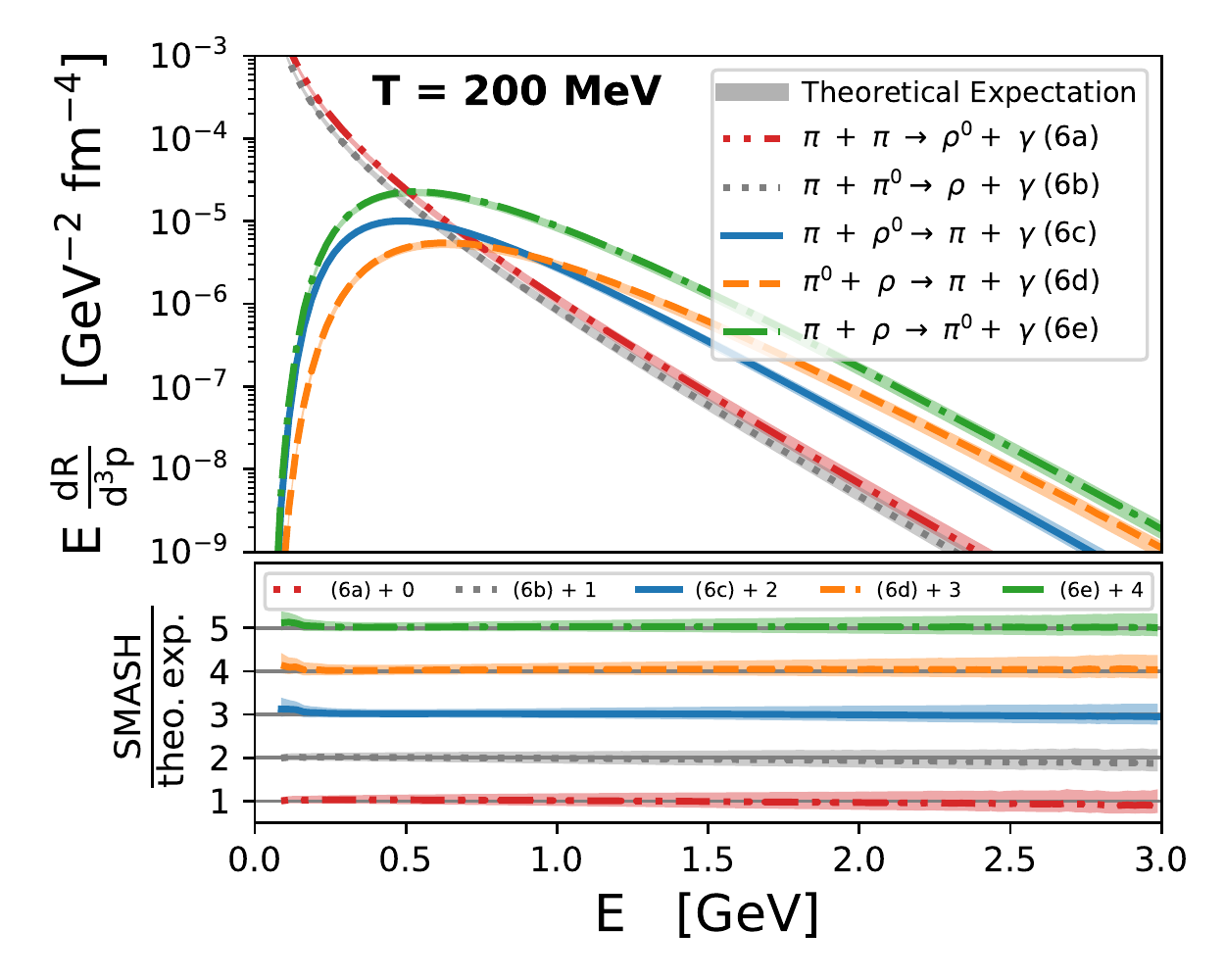}
  	\includegraphics[width=0.45\textwidth]{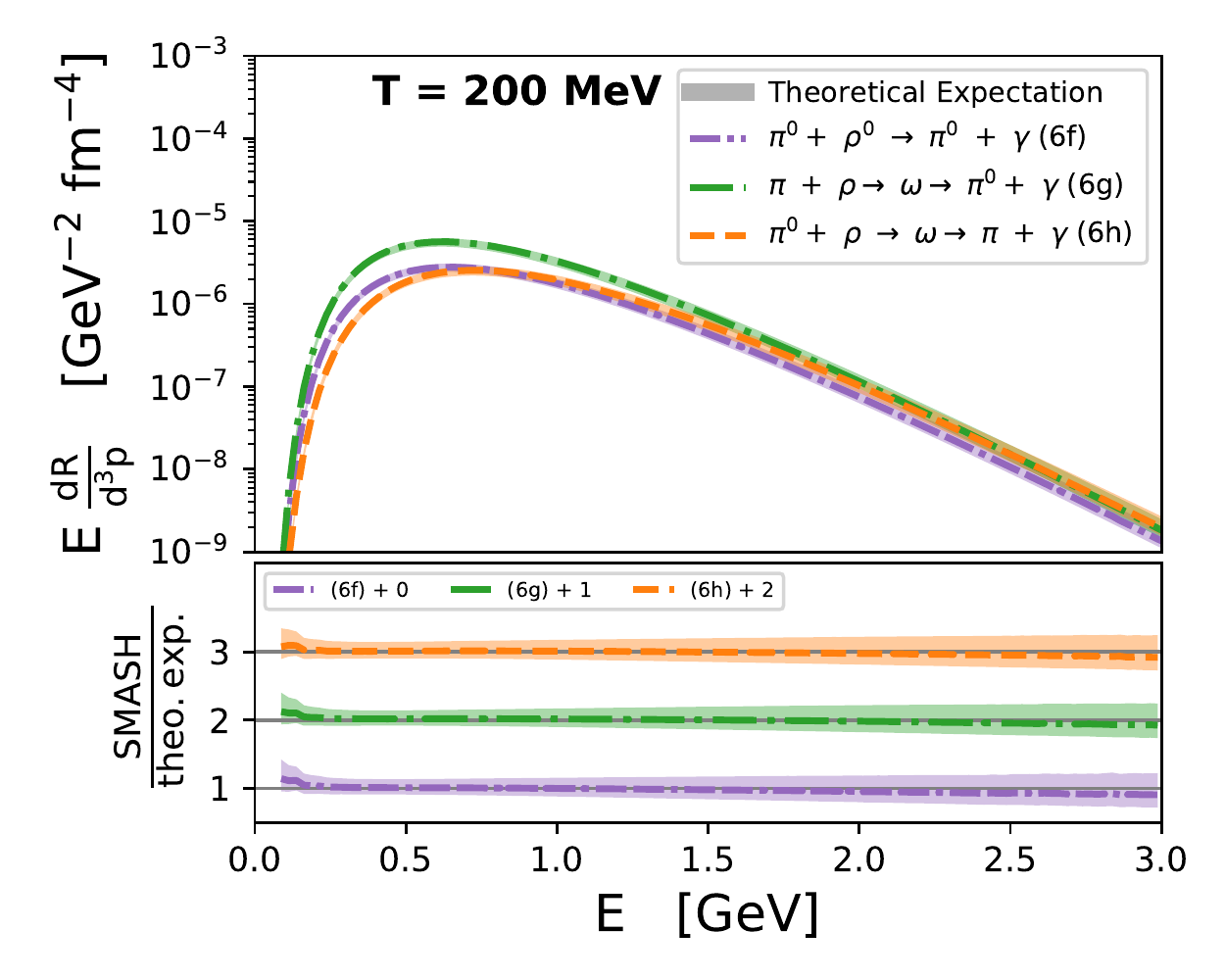}
  	\caption{Thermal photon rates as from SMASH (lines) in comparison to theoretical expectations (bands) at a temperature of $T =$ 200 MeV. See Sec. \ref{Rates} for further details.}
  	\label{Theory_Comp_200}
  \end{figure}

In previous works, thermal photon production in hadronic processes was usually investigated at a temperature of $T = $ 200 MeV. In the above considerations however, all computations are evaluated at $T = $ 150 MeV, as strongly interacting matter is believed to exist in quarks and gluons instead of hadrons at $T = $ 200 MeV. For the sake of completeness, the corresponding results at $T =$ 200 MeV are presented in this section. Qualitatively the results presented in Fig. \ref{Theory_Comp_200} are identical to those at $T = $ 150 MeV in Sec. \ref{ThermalRate}.
  
  \section{\label{app:Broad_Rho}  Current Conservation and Broad $\rho$}
As described in Sec. \ref{broad}, it is possible to extend the derived photon framework to account for finite-width $\rho$ mesons by means of SMASH. However, this transition entails problems regarding current conservation in those production channels where $\rho$ mesons serve simultaneously as initial and intermediate state particles with different masses. As already described in Sec. \ref{broad}, both masses are equated, to circumvent this problem and enforce current conservation in exchange for a physics-wise less complete description. A systematic error is thus introduced into the presented model. To assess the magnitude of the introduced uncertainty, two further analyses are undertaken.
First, the average value of $\Delta$ (Eq. (\ref{Delta})), the term that breaks current conservation once the incoming and the intermediate $\rho$ meson masses are not identical, is investigated. \\
    \begin{figure}
  	\includegraphics[width=0.45\textwidth]{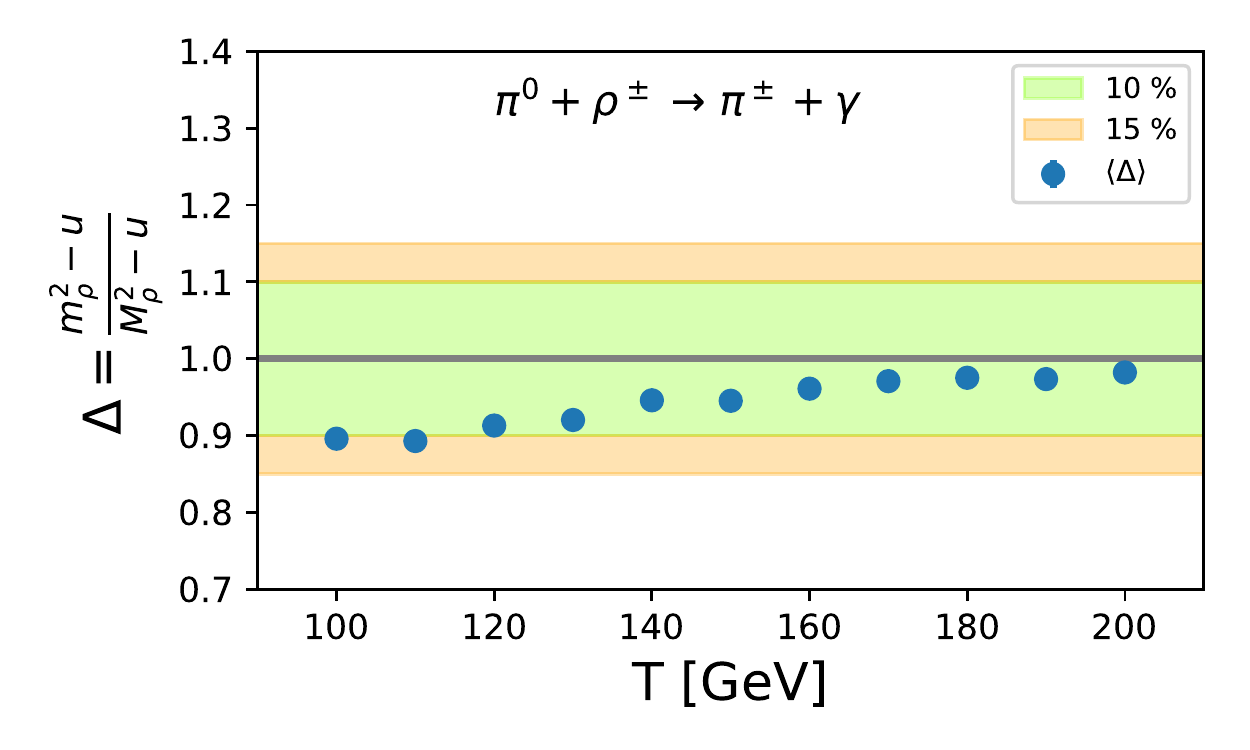}
  	\caption{Average value of the current conservation breaking term, $\Delta$, as a function of the temperature.  $M_\rho = 0.776$ GeV is assumed while $m_\rho$ and $u$ are extracted from SMASH.}
  	\label{Delta_plot}
  \end{figure}
  Fig.~\ref{Delta_plot} shows the average value of $\Delta =  \frac{m_\rho^2 - u}{M_\rho^2 - u}$ for different initialization temperatures of a thermally equilibrated box. Resulting from the perturbative photon treatment in SMASH, the intermediate $\rho$ meson is never actually formed, so its mass is not accessible. To nevertheless estimate the effect of a broad initial $\rho$ meson on $\Delta$, $M_\rho$ is approximated with the $\rho$ pole mass, such that $M_\rho = 0.776$ GeV. The mass of the incoming $\rho$ meson follows from the underlying dynamics in SMASH. It can be observed that the average value of $\Delta$ differs by at most \mbox{11 \%} from the current conserving expectation of $\Delta = 1$ in the temperature range from 100 - 200 MeV. It can further be stated that $\Delta$ approaches unity for rising temperatures. \\
Second, a contact term is derived to explicitly restore current conservation. The cross sections are recalculated, considering the additional (incoherently added) contribution of the contact term, and the resulting photon rates are finally compared to those without considering the additional contribution. This effort is undertaken in the example of process (\ref{C12}) for which current conservation is found to be violated in the case of $m_\rho \neq M_\rho$, since the condition
\begin{align}
k_\mu \mathcal{M}^\mu = 0,
\label{gauge_inv_condition}
\end{align}
is not fulfilled. Here, $k_\mu$ is the photon momentum and $\mathcal{M}^\mu$ the matrix element without the photon polarization vector. The introduced contact term $\mathcal{M}_c$ modifies the matrix element such that
\begin{align}
k_\mu \left(\mathcal{M}^\mu + \mathcal{M}_c^\mu\right) = 0,
\end{align}
and current conservation is restored. Note though that there is no unambiguous definition for $\mathcal{M}_c^\mu$ following condition (\ref{gauge_inv_condition}) and it is possible to construct a contact term 
\begin{align}
\mathcal{M}_{c}^{\prime \ \mu} = (\mathcal{M}_c^\mu + A \ k^\mu)
\end{align}
with $A$ being an arbitrary function of the kinematic variables that still fulfills condition (\ref{gauge_inv_condition}), since $k_\mu k^\mu$ always vanishes. In this assessment, the simplest case is considered with a minimal modification such that $A = 0$.\\ 
\newline
The modified matrix element $\mathcal{M} + \mathcal{M}_c $ is then used to determine the corrected cross section of process (\ref{C12}). Note though, that the contributions are added incoherently, similarly as for the combination of ($\pi,\rho,a_1$)-mediated and $\omega$-mediated processes. \\
    \begin{figure}
  	\includegraphics[width=0.45\textwidth]{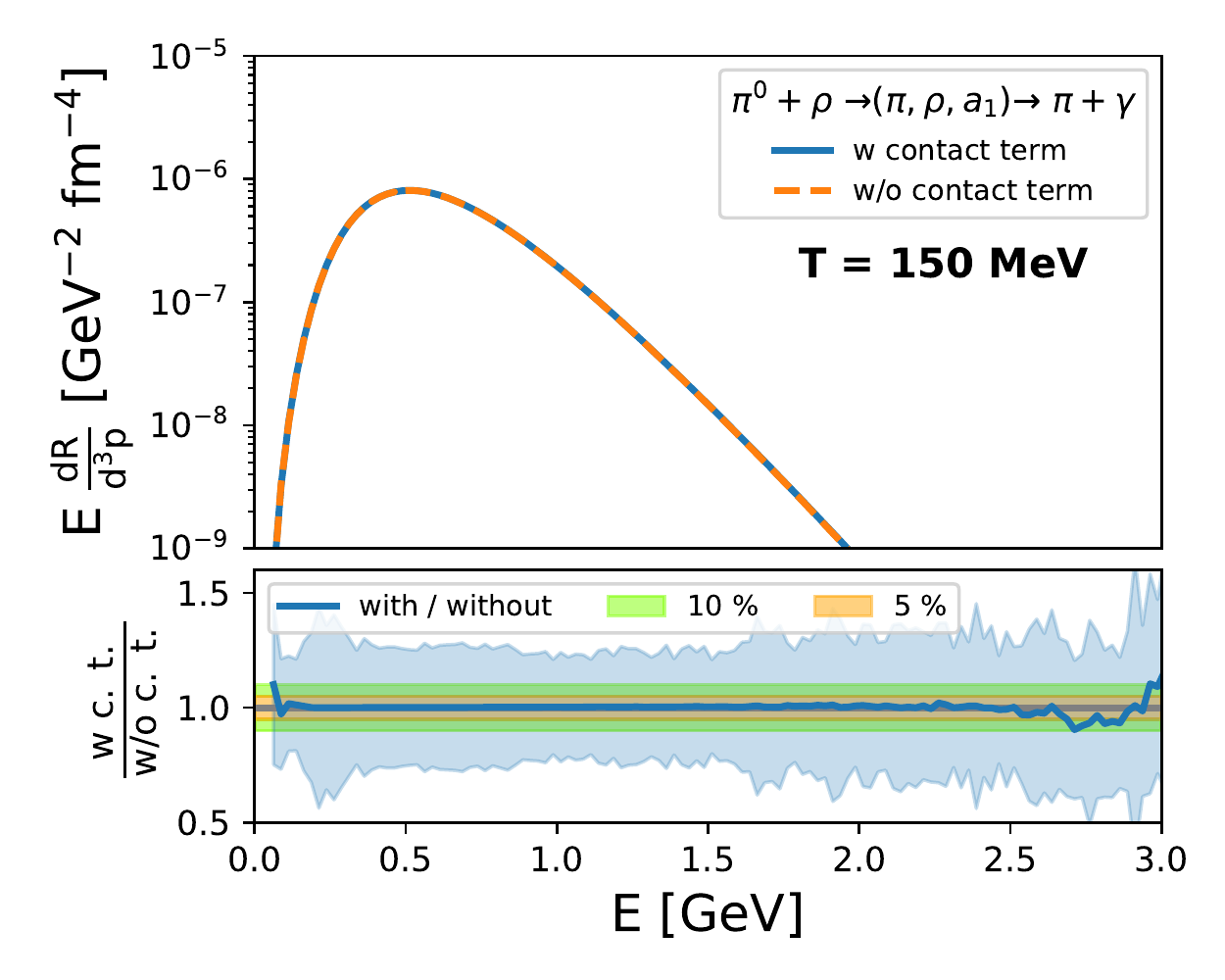}
  	\caption{Thermal photon rate for process (\ref{C12}), $\pi^0 + \rho^\pm \to \pi^\pm + \gamma$, with (solid) and without (dashed) considering an additional contact term to restore current conservation in the case of broad $\rho$ mesons at $T = $ 150 MeV.}
  	\label{Contact_Term_Plot}
  \end{figure} 
Fig.~\ref{Contact_Term_Plot} shows the thermal photon rate for process (\ref{C12}) in the case of broad $\rho$ mesons with (solid line) and without (dashed line) the contribution from the contact term. This additional contribution obviously has only a minor effect on the thermal photon rate. In fact, the ratio of both rates is, within errors, consistent with unity. Generalized to all affected processes, it can be assumed that a consideration of the contact term is not necessary as the photon rate remains unaffected. The treatment of considering broad $\rho$ mesons as described in Sec. \ref{broad} is thus justified.
  
\section{\label{app:param}Parameters}
  For the sake of completeness and reproducibility, the values of all parameters used for the computation of the cross sections depicted in Fig.~\ref{Sigmas} are listed in the following table: \\
  \begin{center}
  \renewcommand{\arraystretch}{1.6}
  \setlength{\tabcolsep}{5pt}
  \begin{tabular}{| l | l |}
  \hline
    $C = 0.059$ &  $\eta_1 = 2.22388$ GeV$^{-1}$\\
    \hline
    $\tilde{g} = 6.4483$ &  $\eta_2 = 2.39014$ GeV$^{-1}$ \\
    \hline
    $\gamma = -0.2913$  &  $m_0 = 0.875$ GeV  \\
      \hline
    $\xi = 0.0585$ & $C_4 = -0.140942$ GeV$^{-2}$ \\
    \hline
   $Z = 0.8429$ & $\Gamma_{a_1} = 0.4$ GeV \\
   \hline
    $\delta = -0.64251$ & $g_{\pi\rho\omega} = \left\{\begin{array}{ll} 11.93$ GeV$^{-1}$ w/o FF  $ \\ 22.6$ GeV$^{-1}$ w FF $ \end{array}\right.$ \\[+0.5cm]
    \hline
  \end{tabular}
  \end{center}
  
\bibliography{Photons}

\end{document}